\documentclass[10pt,twocolumn,twoside]{IEEEtran}    % Enable this line and disable the
\usepackage{amsmath}

\usepackage{amsfonts}

\usepackage{mathtools}

\usepackage{bbold}

\usepackage{upgreek}

\usepackage{graphicx}

\usepackage{subfigure}

\usepackage{bm}

\usepackage{cite}

\usepackage{color}

\usepackage{CJK}

\usepackage{verbatim}

\usepackage{url}

\usepackage{multirow}

\usepackage{array}

\usepackage{booktabs}

\usepackage{colortbl}

\usepackage{algorithm}

\usepackage{algorithmicx,algpseudocode}

\usepackage{diagbox}

\usepackage{makecell}

\usepackage{enumerate}

\usepackage{cases}

\usepackage{lipsum,cuted,arydshln}

\usepackage{threeparttable}

\usepackage{hyperref}
\hypersetup{colorlinks,citecolor=black,
            linkcolor=black,
            urlcolor=black}

\usepackage{varioref}

\usepackage{tablefootnote}

\usepackage[misc]{ifsym}

\usepackage{stfloats}

\usepackage{balance}

\usepackage{caption}

\usepackage[table]{xcolor}

\usepackage{setspace}

\usepackage{amssymb}
\usepackage{nomencl}
\makenomenclature

\usepackage{etoolbox}
\renewcommand\nomgroup[1]{%
  \item[\bfseries
  \ifstrequal{#1}{A}{Sets and Indices}{%
  \ifstrequal{#1}{B}{Parameters}{%
  \ifstrequal{#1}{C}{Variables}{}}}%
]}

\usepackage{pifont}

\def\QEDclosed{\mbox{\rule[0pt]{1.3ex}{1.3ex}}} % ???????
\usepackage[thmmarks]{ntheorem}

\usepackage{nomencl}

\usepackage{wasysym}

\usepackage{hyperref}

\usepackage{algorithm}
\usepackage{algpseudocode}

\makeatletter
\def\UrlAlphabet{%
      \do\a\do\b\do\c\do\d\do\e\do\f\do\g\do\h\do\i\do\j%
      \do\k\do\l\do\m\do\n\do\o\do\p\do\q\do\r\do\s\do\t%
      \do\u\do\v\do\w\do\x\do\y\do\z\do\A\do\B\do\C\do\D%
      \do\E\do\F\do\G\do\H\do\I\do\J\do\K\do\L\do\M\do\N%
      \do\O\do\P\do\Q\do\R\do\S\do\T\do\U\do\V\do\W\do\X%
      \do\Y\do\Z}
\def\UrlDigits{\do\1\do\2\do\3\do\4\do\5\do\6\do\7\do\8\do\9\do\0}
\g@addto@macro{\UrlBreaks}{\UrlOrds}
\g@addto@macro{\UrlBreaks}{\UrlAlphabet}
\g@addto@macro{\UrlBreaks}{\UrlDigits}
\makeatother

{
  \theoremstyle{nonumberplain}
  \theoremheaderfont{\bfseries\em}
  \theorembodyfont{\normalfont}
  \theoremseparator{:}
  \theoremsymbol{\mbox{$\QEDclosed$}}
  
}

\theoremheaderfont{\bf\em}
\theorembodyfont{\normalfont}
\theoremseparator{:}

\newcommand* \cb[1]{\bm{#1}}

\definecolor{gray1}{rgb}{0.9,0.9,0.9}
\definecolor{gray2}{rgb}{0.8,0.8,0.8}
\definecolor{gray3}{rgb}{0.7,0.7,0.7}
\definecolor{white}{rgb}{1,1,1}

\begin{document}


\title{Cyber Recovery from Dynamic Load Altering Attacks: Linking Electricity, Transportation, and Cyber Networks}

\author{
Mengxiang Liu, Zhongda Chu, and Fei Teng
% Mengxiang Liu, Zhenyong Zhang,~\IEEEmembership{Member,~IEEE}, Pudong Ge,~\IEEEmembership{Student~Member,~IEEE},\\ Ruilong Deng,~\IEEEmembership{Senior~Member,~IEEE}, Mingyang Sun,~\IEEEmembership{Member,~IEEE},\\ Jiming Chen,~\IEEEmembership{Fellow,~IEEE}, and Fei Teng,~\IEEEmembership{Senior~Member,~IEEE}
\vspace{-30pt}
% \thanks{{\color{black}This work was supported in part by the National Natural Science Foundation of China under Grants 61833015, 62073285, 62061130220, 61903328, 62103367, in part by the Zhejiang Provincial Natural Science Foundation under Grants LZ21F020006 and LZ22F030010, in part by the Fundamental Research Funds for the Central Universities (226-2022-00120), and in part by the Key Laboratory of Collaborative Sensing and Autonomous Unmanned Systems of Zhejiang Province. (\emph{Corresponding author: Ruilong Deng.})}}
\thanks{The authors are with the Department of Electrical and Electronic Engineering, Imperial College London, London, UK.}
% \thanks{Z. Zhang, R. Deng, M. Sun, and J. Chen are with the State Key Laboratory of Industrial Control Technology and the College of Control Science and Engineering, Zhejiang University, Hangzhou 310027, China (e-mails: \{zhangzhenyong, dengruilong, mingyangsun, cjm\}@zju.edu.cn).}
}
\maketitle
% \vspace{-40pt}
\begin{spacing}{0.88}
\begin{abstract}
To address the increasing vulnerability of power grids, significant attention has been focused on the attack detection and impact mitigation. However, it is still unclear how to effectively and quickly recover the cyber and physical networks from a cyberattack. In this context, this paper presents the first investigation of the Cyber Recovery from Dynamic load altering Attack (CRDA). Considering the interconnection among electricity, transportation, and cyber networks, two essential sub-tasks are formulated for the CRDA: i) Optimal design of repair crew routes to remove installed malware and ii) Adaptive adjustment of system operation to eliminate the mitigation costs while guaranteeing stability. To achieve this, linear stability constraints are obtained by estimating the related eigenvalues under the variation of multiple IBR droop gains based on the sensitivity information of strategically selected sampling points. Moreover, to obtain the robust recovery strategy, the potential counter-measures from the adversary during the recovery process are modeled as maximizing the attack impact of remaining compromised resources in each step. A Mixed-Integer Linear Programming (MILP) problem can be finally formulated for the CRDA with the primary objective to reset involved droop gains and secondarily to repair all compromised loads. Case studies are performed in the modified IEEE 39-bus power system to illustrate the effectiveness of the proposed CRDA compared to the benchmark case.

\end{abstract}

% \mbox{}
% A: Sets and Indices
% B: Parameters
% C: Variables

% \nomenclature[A, 02]{\(c\)}{Speed of light in a vacuum}
% \nomenclature[A, 03]{\(h\)}{Planck constant}
% \nomenclature[A, 01]{\(G\)}{Gravitational constant}
% \nomenclature[B, 03]{\(\mathbb{R}\)}{Real numbers}
% \nomenclature[B, 02]{\(\mathbb{C}\)}{Complex numbers}
% \nomenclature[B, 01]{\(\mathbb{H}\)}{Octonions}

\nomenclature[A, 01]{\(\mathcal{G}\)}{Set of generator buses}
\nomenclature[A, 02]{\(\mathcal{L}\)}{Set of load buses}
\nomenclature[A, 03]{\(\mathcal{A}\)}{Set of compromised load buses and repair depots}
\nomenclature[A, 03]{\(\mathcal{D}\)}{Set of defense load buses with IBRs}
\nomenclature[A, 05]{\(\mathcal{T}\)}{Set of time steps for recovery planning}
\nomenclature[A, 06]{\(\mathcal{C}\)}{Set of repair crews}
\nomenclature[A, 07]{\(\Lambda\)}{Set of eigenvalues of matrix \(\mathcal{B}\)}
\nomenclature[A, 08]{\(\mathcal{M}\)}{Set of combinations of compromised buses' availability along with the recovery process}
\nomenclature[A, 09]{\(\mathcal{J}\)}{Set of combinations of sampling points indexing IBR droop gains}
\nomenclature[A, 10]{\(st/en\)}{Indices of the start/end depots}

\nomenclature[B,02]{\(\overrightarrow{1}/\overrightarrow{0}\)}{Vectors with all one/zero entries}
\nomenclature[B,01]{\(I\)}{Identity matrix}
\nomenclature[B,02]{\(\overleftrightarrow{0}\)}{Matrix with all zero entries}
\nomenclature[B,02]{\(M^{(i)}\)}{Inertial parameter of bus $i\in\mathcal{G}$}
\nomenclature[B,03]{\(D^{(i)}_L/D^{(i)}_G\)}{Damping parameter of bus \(i\in\mathcal{L}/i\in\mathcal{G}\)}
\nomenclature[B,04]{\(K^{(i)}_P/K^{(i)}_I\)}{Proportional/Integral gain of turbine-governor and load-frequency controllers at bus $i\in\mathcal{G}$}
\nomenclature[B,06]{\(P^{(i)}_C\)}{Power output of IBRs at bus \(i\in\mathcal{L}\)}
\nomenclature[B,07]{\(P^{(i)}_L\)}{Power load at bus \(i\in\mathcal{L}\)}
\nomenclature[B,08]{\(P^{(i)}_{LS}/P^{(i)}_{LV}\)}{Secure/Vulnerable power load part at bus \(i\in\mathcal{L}\)}
\nomenclature[B,12]{\(\overline{P}^{(i)}_{LV}\)}{Maximum vulnerable power load at bus \(i\in\mathcal{L}\)}
\nomenclature[B,13]{\(\omega_s^{max}\)}{Safety limitation for generator tripping of bus \(s\in\mathcal{G}\)}
\nomenclature[B,14]{\(\alpha_s^{max}\)}{Safety limitation for load shedding at bus \(s\in\mathcal{L}\)}
\nomenclature[B,15]{\(P^{(i)}_{C*}\)}{Active power reference of the IBRs at bus \(i\in\mathcal{L}\)}
\nomenclature[B,18]{\(P^{(i)}_{C,max}\)}{Maximum active power from IBRs at bus \(i\in\mathcal{L}\)}
% \nomenclature[B,19]{\(\gamma_i\)}{Weight priority of malware removal at bus \(i\in\mathcal{A}\)}
\nomenclature[B,20]{\(\beta_i\)}{Weight parameter of recovering the compromised load bus \(i\in\mathcal{A}\)}
\nomenclature[B,21]{\(B\)}{Large penalty parameter of Big-M method}
\nomenclature[B,22]{\(\epsilon\)}{Small positive parameter}
\nomenclature[B,23]{\(T^{(i,j,c)}\)}{Travel time for crew \(c\) from compromised buses \(i\) to \(j\)}
\nomenclature[B,24]{\(R^{(i,c)}\)}{Repair time of compromised bus \(i\) for crew \(c\)}
\nomenclature[B,24]{\({\cb{k}}_{LG}^{m^*}\)}{Optimal attack control gain vector in scenario \(m\in\mathcal{M}\)}
\nomenclature[B,25]{\(O\)}{Order matrix that sorts the combinations of sampling points indexing IBR droop control gains}
\nomenclature[B,26]{\(\tilde{\cb{k}}_{LG}^{j}\)}{IBR droop control gain vector corresponding to the \(j\)-th column of \(O\)}
\nomenclature[B,27]{\(\lambda_{n,0}^{(j,m^*)}\)}{Start eigenvalue at sampling point \(\tilde{\cb{k}}_{LG}^j\) used for the estimation of \(\lambda_n\) under attack parameter \({\cb{k}}_{LG}^{m^*}\)}
\nomenclature[B,28]{\(\tilde{\cb{k}}_{LG,n,0}^{(j,m^*)}\)}{Start droop gain vector at sampling point \(\tilde{\cb{k}}_{LG}^j\) used for the estimation of \(\lambda_n\) under attack parameter \({\cb{k}}_{LG}^{m^*}\)}
\nomenclature[B,29]{\(\frac{\partial\lambda_n}{\partial\tilde{\cb{k}}_{LG}}\vert_{(j,m^*)}\)}{Eigenvalue sensitivity at sampling point \(\tilde{\cb{k}}_{LG}^j\) used for the estimation of \(\lambda_n\) under attack parameter \({\cb{k}}_{LG}^{m^*}\)}
\nomenclature[B,29]{\(\lambda_{n}^{(j,m^*)}\)}{Matrix \(\mathcal{B}\)'s \(n\)-th eigenvalue under IBR droop gain \(\tilde{\cb{k}}_{LG}^j\) and attack parameter \({\cb{k}}_{LG}^{m^*}\)}
\nomenclature[B,30]{\(\cb{r}_n^{(j,m^*)}/\cb{l}_n^{(j,m^*)}\)}{Right/Left eigenvector of \(\lambda_n\) under IBR droop gain \(\tilde{\cb{k}}_{LG}^j\) and attack parameter \({\cb{k}}_{LG}^{m^*}\)}
\nomenclature[B,31]{\(\widetilde{K}_{LG}^{(i,s,l)}\)}{IBR droop gain of bus \(i\in\mathcal{D}\) with frequency measurement from bus \(s\in\mathcal{G}\) at the \(l\)-th sampling point}
\nomenclature[B,31]{\(L^{(i,l)}/U^{(i,l)}\)}{Lower/Upper bound of the \(l\)-th range segment describing the IBR droop gain of bus \(i\in\mathcal{D}\)}

\nomenclature[C,01]{\(\delta_i\)}{Phase angle of bus \(i\in\mathcal{G}\)}
\nomenclature[C,02]{\(\omega_i\)}{Rotor frequency deviation of bus \(i\in\mathcal{G}\)}
\nomenclature[C,03]{\(\theta_i\)}{Phase angle of bus \(i\in\mathcal{L}\)}
\nomenclature[C,04]{\(\alpha_i\)}{Frequency deviation of bus \(i\in\mathcal{L}\)}
\nomenclature[C,05]{\(K^{(i,s)}_{LG}/K^{(i,s)}_{LL}\)}{Attack control gain at bus \(i\in\mathcal{L}\) with frequency measurement from bus \(s\in\mathcal{G}\)/\(\mathcal{L}\) }
\nomenclature[C,06]{\(\widetilde{K}^{(i,s)}_{LG}/\widetilde{K}^{(i,s)}_{LL}\)}{Droop control gain at bus \(i\in\mathcal{L}\) with frequency measurement from bus \(s\in\mathcal{G}\)/\(\mathcal{L}\)}
\nomenclature[C,07]{\(X_{RC}^{(i,j,c)}\)}{Binary variable indicating if crew \(c\) moves from buses \(i\) to \(j \in \mathcal{A}\)}
\nomenclature[C,08]{\(X_{RC^*}^{(i,j,c)}\)}{Product of binary variables \(X_{RC}^{(i,j,c)}\) and \(X_{RC}^{(j,i,c)}\)}
% \nomenclature[C,08]{\(x_{i,j}^{c*}\)}{Binary variable indicating the product of \(x_{i,j}^{c}\) and \(x_{j,i}^{c}\)}
\nomenclature[C,08]{\(Y_{RC}^{(i,c)}\)}{Binary variable indicating if crew \(c\) visited bus \(i\in\mathcal{A}\)}
\nomenclature[C,09]{\(AT_{RC}^{(i,c)}\)}{Arrival time of crew \(c\) at bus \(i\in\mathcal{A}\)}
\nomenclature[C,10]{\(F_{RC}^{(i,t)}\)}{Binary variable indicating if bus \(i\in\mathcal{A}\) is repaired at time \(t\in\mathcal{T}\)}
\nomenclature[C,11]{\(Z_{RC}^{(i,t)}\)}{Binary variable indicating if bus \(i\in\mathcal{A}\) is available at time \(t\in\mathcal{T}\)}
\nomenclature[C,12]{\(S_{DG}^{(m,t)}\)}{Binary variable indicating if sensitivity matrices of \(m\in\mathcal{M}\)-th scenario is selected at time \(t\in\mathcal{T}\)}
\nomenclature[C,14]{\(\lambda_{0}^{(n,j,t)}\)}{Start eigenvalue at sampling point \(\tilde{\cb{k}}_{LG}^j\) used for the estimation \(\lambda_n\) at time \(t\in\mathcal{T}\)}
\nomenclature[C,15]{\(\tilde{\cb{k}}_{LG,0}^{(n,j,t)}\)}{Start droop gain vector at sampling point \(\tilde{\cb{k}}_{LG}^j\) used for the estimation \(\lambda_n\) at time \(t\in\mathcal{T}\)}
\nomenclature[C,16]{\(\frac{\partial\lambda_n}{\partial\tilde{\cb{k}}_{LG}}\vert_{(j,t)}\)}{Eigenvalue sensitivity at sampling point \(\tilde{\cb{k}}_{LG}^j\) used for the estimation of \(\lambda_n\) at time \(t\in\mathcal{T}\)}
\nomenclature[C,17]{\(T_1^{(i,l,t)}/T_2^{(i,l,t)}\)}{Binary variable indicating if \(\widetilde{K}_{LG}^{(i,s,t)}\) is larger/smaller than the \(l\)-th lower/upper bound at time \(t\in\mathcal{T}\)}
\nomenclature[C,18]{\(T^{(i,l,t)}\)}{Binary variable indicating if \(\widetilde{K}_{LG}^{(i,s,t)}\) is within the \(l\)-th range segment at time \(t\in\mathcal{T}\)}
\nomenclature[C,19]{\(\Psi^{(j,t)}\)}{Binary variable indicating if the IBR droop gains match sampling point combination \(j\in\mathcal{J}\) at time \(t\in\mathcal{T}\)}
\nomenclature[C,20]{\(\tilde{\cb{k}}_{LG}^{(t)}\)}{IBR droop gain vector comprising \(\widetilde{K}_{LG}^{(i,s,t)}, \forall i \in\mathcal{D}\) at time \(t\in\mathcal{T}\)}
\nomenclature[C,21]{\(\hat{\lambda}_{n,t}\)}{Estimation of the \(n\)-th eigenvalue at time \(t\)}

\printnomenclature[1.8cm]

% \vspace{-10pt}
\section{Introduction}
The Internet-of-Things (IoT) has been recognized a key enabling technology for smart grid benefiting from its wide connectivity over everyday devices \cite{7879243,8449080,7467406}. Among the widespread IoT-enabling applications, smart home is one of the most typical scenarios in which home appliances, including lighting, heating, security, and refrigeration systems, can be controlled remotely and via the Internet by mobile devices or by digital assistants such as Amazon’s Alexa \cite{lopatovska2019talk}. Although these advanced IoT technologies can significantly improve the efficiency of home energy management and provide grid-support services, numerous attack surfaces are also exposed by the adopted public Internet services. Studies have demonstrated that IoT devices from cameras to locks can be compromised either
directly or through their designated mobile applications
by an adversary \cite{fernandes2016security}. Besides the consequences of IoT vulnerabilities on personal privacy and security, the collective effect of these vulnerabilities was also demonstrated by the Mirai botnet, comprising six hundred thousand compromised devices targeting
victim servers \cite{antonakakis2017understanding}.

Once a large enough number of high-wattage appliances like air conditioners and water heaters are compromised, the adversary is able to disrupt the normal power grid operations by manipulating these power demands. The concept of this kind of attack was first proposed by Mohsenian \textit{et al.} in \cite{mohsenian2011distributed} and was termed as static load altering attack (SLAA). Following this, Soltan \textit{et al.} verified the feasibility of launching SLAAs with compromised IoT appliances, and demonstrated that line failures, economic losses, and frequency instability can be induced by synchronously switching on/off all compromised IoT devices \cite{soltan2018blackiot}. 
% Moreover, the potential of adopting load temporal fluctuations and intermittent renewable power outputs in executing LAAs was investigated in \cite{ospina2020feasibility,lakshminarayana2022load}. 
On top of SLAA, Amini \textit{et al.} proposed a variant, named as dynamic load altering attack (DLAA), where the load manipulation is controlled based on a feedback from the power system frequency \cite{amini2016dynamic}. Compared with the SLAA, the later proposed DLAA requires extra frequency measurements and system knowledge, but it can destabilize the frequency control loop much more easily following appropriate attack strategies. The frequency acquisition does not remain challenging as the adversary can obtain the required data from Phasor Measurement Units (PMUs) and frequency-responsive loads. Recently, the least-effort destabilizing DLAA was identified using the theory of second-order dynamical systems \cite{lakshminarayana2021analysis}.

Numerous countermeasures have been proposed against SLAAs and DLAAs, and are divided into prevention, detection, and mitigation methods according to the occurrence time of attacks \cite{liu2023enhancing}. At the pre-attack stage, both prevention strategies from information technology (IT) and operation technology (OT) sides can be deployed. In the IT domain, encryption and authorization technologies can be employed to prevent the adversary from intruding into the home area network and compromising IoT appliances. Amini \textit{et al.} designed the optimal load protection scheme under which the DLAA can never destabilize the frequency dynamics \cite{amini2016dynamic}. In the OT domain, the generator operating points can be adjusted such that no power line will be overloaded instantly after any potential SLAAs \cite{8734744}. Furthermore, a model-free frequency controller was designed to tolerate various SLAA scenarios via the pre-trained off-line strategy \cite{chen2020load}. These prevention strategies are usually designed offline considering the worst attack case and thus easily suffer from cost-inefficiency related issues \cite{7210191,9497752}. Besides, it is difficult to prevent all intrusions due to unpredictable zero-day vulnerabilities and inexhaustible attack scenarios, implying the criticality of the subsequent response strategies.

During the attack stage, it is necessary to identify attack locations and infer attack parameters in a timely manner. The initial attempt to detect DLAAs was accomplished by matching attack signatures of system poles based on frequency-domain smart meter readings \cite{7436350}, which was then extended to identify attack locations \cite{8240622}. To address the time-consuming issue of the frequency-domain enabling 
attack perception methods, the unscented Kalman filter was adopted to perform dual state estimation to identify the attack locations \cite{7963476}. However, the above methods cannot be applied trivially to practical large-scale system with nonlinear system dynamics due to their limited scalability. Towards this end, physics-informed data-driven methods were proposed to identify attack locations and estimate attack parameters using real-time frequency/phase angle measurements monitored by PMUs \cite{lakshminarayana2022data}. The following action after knowing the attack locations and associated parameters is to mitigate the adverse attack impact. Based on the online attack identification information, Chen \textit{et al.} proposed an adaptive mitigation scheme against SLAAs where the candidate strategy is selected from the off-line trained strategy pool after scenario matching \cite{chen2020load}. By leveraging the flexibility of inverter-based resources (IBRs), a Cyber-Resilient Economic Dispatch (CRED) scheme was proposed to mitigate the destabilizing impact of SLAAs and DLAAs \cite{chu2022mitigating}. In particular, CRED coordinates the frequency droop control gains of IBRs in the system to mitigate the adverse impact while minimizing the overall operational cost.

\begin{figure}
\includegraphics[width=9cm]{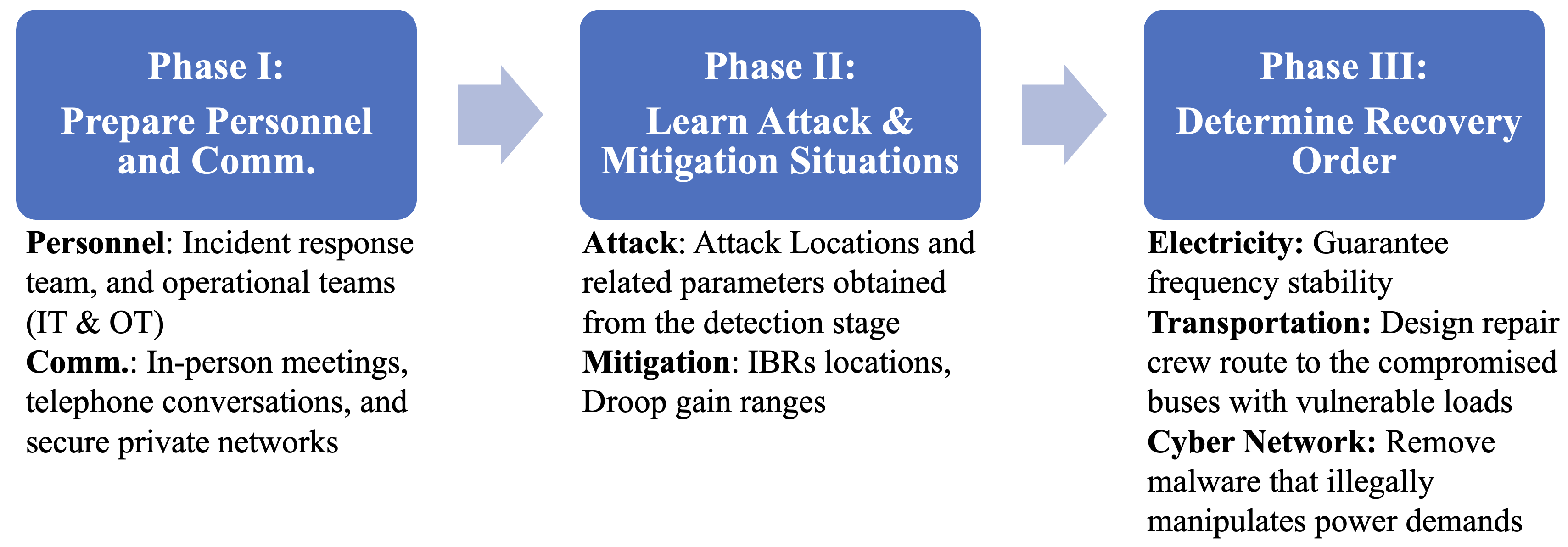}
\captionsetup{font={small}}
{\caption{Process of cyber recovery from DLAAs}\label{fig:recover process}}
\end{figure}

However, for a holistic cyber-resiliency-enhancement framework \cite{liu2023enhancing}, the mitigation stage will not be the end as the reconfigured control algorithm has performance degradation \cite{chen2020load} and the CRED scheme requires extra generation costs \cite{chu2022mitigating}. Hence, in the last-but-not-least post-attack stage, recovery methods that remove malware from the compromised IoT appliances and restore the OT side's control and operation schemes are vital but have been paid little attention. According to the NIST's guide for cybersecurity event recovery \cite{bartock2016guide}, the process of Cyber Recovery from DLAAs (CRDA) is customized in Fig. \ref{fig:recover process}, mainly including three phases: i) Prepare recovery personnel and determine secure communication ways, ii) Identify current attack and mitigation situations, and iii) Determine the recovery order considering the links among electricity, transportation, and cyber networks. The first two phases have been covered by existing prevention and detection related literature, while the critical Phase iii that determines the recovery order has not been addressed yet. We have to note here that the considered Cyber Recovery from DLAA (CRDA) is similar to the physical recovery under natural disasters \cite{10050344,arif2017power}, but there are essential differences between them: i) The compromised loads are still connected to the power grid during the recovery stage; ii) The operational strategy, such as IBR droop gains, need to be adjusted online in response to the repair of compromised loads; iii) The adversary may alter the attack parameters when perceiving the recovery actions;

To fill this gap, this paper targets the crucial recovery order determination problem in the CRDA, assuming that the state-of-the-art mitigation strategy, i.e., CRED \cite{chu2022mitigating}, has been activated temporarily to eliminate the destabilizing attack impact. Essentially, the proposed CRDA needs to accomplish two sub-tasks: i) Optimal route of repair crews to remove the malware installed at compromised loads, and ii) Adaptive adjustment of IBR droop gains to alleviate the cost of CRED while guaranteeing stability. Accordingly, the primary objective of CPRP is to eliminate the IBR usage for impact mitigation, i.e., reset associated droop gains to their previous values, while the second objective is to repair all compromised loads. Finally, the CRDA will be formulated as a Mixed-Integer Linear Programming (MILP) problem. The original contributions of this paper are summarized as:
\begin{itemize}
    \item The CRDA is investigated for the first time and formulated as an MILP considering Repair Crew Route (RCR) and adaptive Droop Gain Adjustment Stability (DGAS) constraints.
    \item To formulate the linear stability constraints, sensitivity information of multiple sampling points are strategically selected to estimate the related eigenvalues where more than one aggregated IBR droop gains could vary.
    \item The robust recovery strategy is obtained by incorporating the worst-case attack movements based on the available compromised loads in each recovery step into the MILP problem.
\end{itemize}

The remainders of this paper are listed as follows: Section \ref{Sec: System model} introduces the frequency dynamic model, attack model, and CRED scheme. Section \ref{Sec: Problem Formulation} formulates the MILP problem for the CRDA, and Section \ref{Sec: RCR constrains} and Section \ref{Sec: DGA constraints} detail the RCR and DGAS constraints, respectively. Section \ref{section: case studies} demonstrates the results of case studies and Section \ref{Sec: Summary} summarizes this paper and indicates future directions.

\section{System Models and Cyber-Resilient Mitigation}\label{Sec: System model}
In this section, we introduce the frequency dynamic model, attack model, and CRED scheme.

\subsection{Frequency Dynamic Model}
We consider a power system consisting of a set of $\mathcal{N}=\mathcal{G}\cup\mathcal{L}$ buses. The number of buses $|\mathcal{N}|=|\mathcal{G}|+|\mathcal{L}|$, where the symbol $|\cdot|$ calculates the size of a set. The linearized frequency dynamic model can be described by the following set of differential equations \cite{glover2012power}:
\begin{align}\label{eq:frequency dynamics}
&\begin{bmatrix}
I & \overleftrightarrow{0} & \overleftrightarrow{0} & \overleftrightarrow{0} \\
\overleftrightarrow{0} & I & \overleftrightarrow{0} & \overleftrightarrow{0} \\
\overleftrightarrow{0} & \overleftrightarrow{0} & -M & \overleftrightarrow{0} \\
\overleftrightarrow{0} & \overleftrightarrow{0} & \overleftrightarrow{0} & \overleftrightarrow{0}
\end{bmatrix}
\begin{bmatrix}
\dot{\cb{\delta}}\\
\dot{\cb{\theta}}\\
\dot{\cb{\omega}}\\
\dot{\cb{\alpha}}
\end{bmatrix}=
\begin{bmatrix}
\overrightarrow{0}\\
\overrightarrow{0}\\
\overrightarrow{0}\\
\cb{p}_L-\cb{p}_C
\end{bmatrix}+ \nonumber \\ 
&\ \quad\quad\quad\quad{\small\begin{bmatrix}
\overleftrightarrow{0} & \overleftrightarrow{0} & I & \overleftrightarrow{0} \\
\overleftrightarrow{0} & \overleftrightarrow{0} & \overleftrightarrow{0} & I \\
K_I+H_{GG} & H_{GL} & K_P+D_G & \overleftrightarrow{0} \\
H_{LG} & H_{LL} & \overleftrightarrow{0} & D_L
\end{bmatrix}}
\begin{bmatrix}
\cb{\delta}\\
\cb{\theta}\\
\cb{\omega}\\
\cb{\alpha}
\end{bmatrix},
\end{align}
where vectors $\cb{\delta}, \cb{\omega} \in \mathcal{R}^{|\mathcal{G}|}$ and $\cb{\theta}, \cb{\alpha} \in \mathcal{R}^{|\mathcal{L}|}$ are compacted forms of $\delta_i, \omega_i, i \in\mathcal{G}$, and $\theta_i, \alpha_i, i \in\mathcal{L}$, respectively, and vectors $\cb{p}_L, \cb{p}_C \in \mathcal{R}^{|\mathcal{L}|}$ comprise $P^{(i)}_L, P^{(i)}_C, \forall i \in \mathcal{L}$. Diagonal matrices $M, D_{G} \in \mathcal{R}^{|\mathcal{G}|\times|\mathcal{G}|}$ and $D_L \in \mathcal{R}^{|\mathcal{L}|\times|\mathcal{L}|}$ are established using $M^{(i)}, D^{(i)}_{G}, \forall i 
\in\mathcal{G}$ and $D^{(i)}_L,\forall i 
\in\mathcal{L}$, respectively. Moreover, diagonal matrices $K_P, K_I \in \mathcal{R}^{|\mathcal{G}|\times|\mathcal{G}|}$ collect $K^{(i)}_P$ and $K^{(i)}_I, \forall i \in\mathcal{G}$. Matrices $H_{GG}\in\mathcal{R}^{|\mathcal{G}|\times|\mathcal{G}|}, H_{GL}\in\mathcal{R}^{|\mathcal{G}|\times|\mathcal{L}|}, H_{LG}\in\mathcal{R}^{|\mathcal{L}|\times|\mathcal{G}|}$, and $H_{LL}\in\mathcal{R}^{|\mathcal{L}|\times|\mathcal{L}|}$ are sub-matrices of the admittance matrix derived from the simplified DC power flow model \cite{stott2009dc}.

\subsection{Attack Model}
This subsection will illustrates the DLAA model. The DLAA tries to destabilize the frequency from its nominal value (50 Hz in Europe and 60Hz in North America) by manipulating the power consumption of load buses \cite{soltan2018blackiot}. The adversary is assumed to be able to access an IoT botnet of many high wattage smart appliances like air conditioners within a number of load buses. Since most of IoT appliances are controlled by mobile phone Apps, access to mobile phones or corresponding Apps can also control these appliances. When the number of compromised IoT appliances is large enough, the adversary can potentially achieve malicious objectives by turning on/off these appliances in different locations remotely following specific manners. Consider a single-point DLAA against load bus $i \in \mathcal{L}$, the power consumption can be modeled as $P^{(i)}_L=P^{(i)}_{LS}+P^{(i)}_{LV}$. To maximize the attack impact, the power consumption should be manipulated in a opposite direction with respect to the frequency deviation. Here the frequency measurement may be obtained from PMUs or frequency-sensitive loads \cite{amini2016dynamic}. Since the frequency sensor bus $s$ can be either a generator or load bus, we have
\begin{align}\label{eq: DLAA}
P^{(i)}_{LV}=-K^{(i,s)}_{LG}\omega_s-K^{(i,s)}_{LL}\alpha_s, s \in \mathcal{N}.
\end{align}

As practically the frequency measurement is usually obtained from one sensor, only one of the two attack control gains $K^{(i,s)}_{LG}, K^{(i,s)}_{LL}\ge0$ can be nonzero. By carefully designing these attack control gains, the eigenvalues of the differential equations \eqref{eq:frequency dynamics} can be driven to the unstable area (like making the real part of eigenvalue positive), and thus destabilizing the frequency. Since the compromised power consumption is limited, the attack control gains should be restricted by
\begin{subequations}\label{eq: attack control gain range}
\begin{align}
K^{(i,s)}_{LG}\omega_s^{max} &\le \frac{1}{2}\overline{P}^{(i)}_{LV}, s\in\mathcal{G} \\
K^{(i,s)}_{LL}\alpha_s^{max} &\le \frac{1}{2}\overline{P}^{(i)}_{LV}, s\in\mathcal{L},
\end{align}
\end{subequations}where $\omega_s^{max},\alpha_s^{max}>0$, and the coefficient $\frac{1}{2}$ on the right side of the equation indicates that the power consumption manipulation should allow both under and over frequency fluctuations.
% \vspace{-10pt}

\subsection{Cyber-Resilient Economic Dispatch Scheme}
By real-time monitoring and analyzing the grid’s physical signals, the attack locations that contain compromised loads can be identified timely, with the attack control gains being inferred accurately \cite{lakshminarayana2022data}. Based on the identification and inference results, the CRED scheme was proposed to mitigate the destabilizing impact of DLAAs using the damping provision capability of large-scale penetrated IBRs \cite{chu2022mitigating}. Consider the IBRs connected to bus $i\in\mathcal{L}$, its power output $P^{(j)}_C$ can be calculated as
\begin{align}\label{eq: Cyber-Safe}
P^{(i)}_C=P^{(i)}_{C*}-\widetilde{K}^{(i,s)}_{LG}\omega_s-\widetilde{K}^{(i,s)}_{LL}\alpha_s, s \in \mathcal{N},
\end{align}Similarly, only one of the droop gains $\widetilde{K}^{(i,s)}_{LG},\widetilde{K}^{(i,s)}_{LL}\ge 0$ can be nonzero depending on the sensor type. Given the realistic constraints on IBR power outputs, the droop gains are restricted by
{\small
\begin{subequations}
\begin{align}
P^{(i)}_{C*}+\widetilde{K}^{(i,s)}_{LG}\omega_s^{max} &\le P^{(i)}_{C,max}, P^{(i)}_{C*}+\widetilde{K}^{(i,s)}_{LL}\alpha_s^{max} \le P^{(i)}_{C,max} \label{eq: IBR droop gain ranges1}, \\
P^{(i)}_{C*}-\widetilde{K}^{(i,s)}_{LG}\omega_s^{max} &> 0, 
\quad\quad\quad P^{(i)}_{C*}-\widetilde{K}^{(i,s)}_{LL}\alpha_s^{max} > 0, \label{eq: IBR droop gain ranges2}
\end{align}
\end{subequations}
}

Combine equations \eqref{eq:frequency dynamics}, \eqref{eq: DLAA}, and \eqref{eq: Cyber-Safe}, the frequency dynamics can be rewritten as
\begin{align}\label{eq:frequency dynamics DLAA and Cyber Safe}
% \tiny
&\begin{bmatrix}
% \tiny
I & \overleftrightarrow{0} & \overleftrightarrow{0} & \overleftrightarrow{0} \\
\overleftrightarrow{0} & I & \overleftrightarrow{0} & \overleftrightarrow{0} \\
\overleftrightarrow{0} & \overleftrightarrow{0} & -M & \overleftrightarrow{0} \\
\overleftrightarrow{0} & \overleftrightarrow{0} & \overleftrightarrow{0} & \overleftrightarrow{0}
\end{bmatrix}
\begin{bmatrix}
\dot{\cb{\delta}}\\
\dot{\cb{\theta}}\\
\dot{\cb{\omega}}\\
\dot{\cb{\alpha}}
\end{bmatrix}=
\begin{bmatrix}
\overrightarrow{0}\\
\overrightarrow{0}\\
\overrightarrow{0}\\
\cb{p}_{LS}-\cb{p}_{C*}
\end{bmatrix}+ \\ \nonumber
&{\small\begin{bmatrix}
\overleftrightarrow{0} & \overleftrightarrow{0} & I & \overleftrightarrow{0} \\
\overleftrightarrow{0} & \overleftrightarrow{0} & \overleftrightarrow{0} & I \\
K_I+H_{GG} & H_{GL} & K_P+D_G & \overleftrightarrow{0} \\
H_{LG} & H_{LL} & -K_{LG}+\widetilde{K}_{LG} & D_L-K_{LL}+\widetilde{K}_{LL}
\end{bmatrix}}
{\small\begin{bmatrix}
\cb{\delta}\\
\cb{\theta}\\
\cb{\omega}\\
\cb{\alpha}
\end{bmatrix}},
\end{align}where vectors $\cb{p}_{LS}, \cb{p}_{C*} \in \mathcal{R}^{|\mathcal{L}|}$ include all secure loads and IBR power references, respectively, matrices $K_{LG}=[K^{(i,s)}_{LG}]_{|\mathcal{L}|\times|\mathcal{G}|},\widetilde{K}_{LG}=[\widetilde{K}^{(i,s)}_{LG}]_{|\mathcal{L}|\times|\mathcal{G}|}$ and $K^{LL}=[K^{(i,s)}_{LL}]_{|\mathcal{L}|\times|\mathcal{L}|},\widetilde{K}_{LL}=[\widetilde{K}^{(i,s)}_{LL}]_{|\mathcal{L}|\times|\mathcal{L}|}$ collect the attack control gains and IBR droop gains.

\section{Problem Formulation}\label{Sec: Problem Formulation}
The CRED scheme will only be activated when the DLAA is detected to avoid the frequency destabilization consequence, as the associated IBR droop control requires the deloading of RES and thus will increase the generation cost of synchronous machines. Hence, the CRED should be appropriately integrated into the power grid operation to balance the trade-off between cost and stability \cite{chu2022mitigating}. Another critical yet unsolved issue is how to securely and efficiently quit the CRED scheme and restore to the normal operational mode, which falls in the area of recovery planning under cyberattack events \cite{bartock2016guide}. Assuming that the first two phases demonstrated in Fig. \ref{fig:recover process} have been completed, this paper focuses on the third phase, i.e., how to schedule repair crews to remove the malware compromising IoT-based loads and adjust the IBR droop gains accordingly to restore to the normal operational mode. To formulate the CRDA as an MILP problem, there are mainly three challenges to be addressed: 1) Integrate the RCR and droop gain adjustment tasks into the MILP problem; 2) Model attack movements during the recovery process in the MILP problem; 3) Characterize the eigenvalue variation of dynamical system \eqref{eq:frequency dynamics DLAA and Cyber Safe} under DLAAs, CRED, and recovery actions in the MILP problem.

\subsection{Cyber Recovery from DLAAs}
The CRDA includes two sub-tasks. The first sub-task is the optimal RCR characterized by repair crews, repair resources, compromised load buses, and the transportation paths between them, with variables 
\begin{align*}
\Gamma=\Big\{X_{RC}^{(i,j,c)},X_{RC^*}^{(i,j,c)},Y_{RC}^{(i,c)},AT_{RC}^{(i,c)},F_{RC}^{(i,t)},Z_{RC}^{(i,t)}\Big\}.
\end{align*} The second task is the adaptive DGA {\color{black}to eliminate the extra generation costs while guaranteeing frequency stability} after each malware removal action, with variables 
\begin{align*}
\Upsilon=\Big\{S_{DG}^{(m,t)},\lambda_{0}^{(n,j,t)},\tilde{\cb{k}}_{LG,0}^{(n,j,t)},\frac{\partial \lambda_n}{\partial \tilde{\cb{k}}_{LG}}\Big|_{(n,j,t)},T^{(i,l,t)}_{1},T^{(i,l,t)}_{2},\\
T^{(i,l,t)},\Psi^{(j,t)},\tilde{\cb{k}}_{LG}^{(t)}, \hat{\lambda}^{(n,t)} \Big\}.
\end{align*}

Without loss of generality, we only show the case where $K_{LL}^{(i,s)}=\widetilde{K}_{LL}^{(i,s)}=0,\forall i,s\in\mathcal{L}$ in this paper. The malware removal in the first sub-task will activate the DGA in the second task, and in turn the frequency stability criteria in the second sub-task will guide the RCR design in the first task. Hence, these two tasks are closely coupled and should be jointly addressed in an integrated problem. For the sake of practical usage, the RCR and DGA are both recorded in $|\mathcal{T}|$ discrete time steps. The objective functions and constraints of the proposed CRDA can be thus written as
\begin{subequations}\label{eq: overall objective function}
\begin{align}\label{eq: objective function}
    &\min_{\Gamma,\Upsilon} \sum_{\forall t\in\mathcal{T}}\sum_{\forall i\in\mathcal{D}}\widetilde{K}_{LG}^{(i,s,t)}-\sum_{\forall t\in\mathcal{T}}\sum_{\forall i \in \mathcal{A}/\{st,en\}}\beta_i Z_{RC}^{(i,t)}, \\
    &\text{s.t.\quad\quad RCR constraints: } \eqref{eq: RCR constraints 1}, \eqref{eq: RCR constraint 2}-\eqref{eq: RCR constraint 7}, \\
    &\text{\quad\quad\quad DGAS constraints: } \eqref{eq: DGA constraints1}, \eqref{eq: DGA constraints2}, \eqref{eq: DGA constraints3}-\eqref{eq: DGA constraints6}.
\end{align}
\end{subequations}

Functions \eqref{eq: objective function} include two essential objectives: i) reset all droop gains of IBRs, and ii) remove all malware installed into IoT-based appliances as soon as possible. Objective i) aims to mitigate the economic loss resulted from the deloading of RESs, and thus is given the primary priority. While Objective ii) aims to recapture the control privilege of IoT-based appliances from the adversary, and enable the customer's legitimate access to these appliances through mobile and corresponding Apps. Objective ii) is given the second priority since it mainly recovers the internet services of controlling IoT-based appliances. Hence, the weight parameters $\beta_i>0,\forall i \in \mathcal{A}/\{st,en\}$ are assumed to be far smaller than $1$. In general, these weight parameters are set according to the amount of vulnerable loads at compromised buses, and can be revised by the system operator. The subsequent sections will detail the formulation and derivation of RCR and DGAS constraints.
% \vspace{-30pt}
\section{Repair Crew Route Constraints}\label{Sec: RCR constrains}
The RCR constraints involve the determination of repair crew route, repair statuses and availability of compromised buses over time. In general, these constraints can be divided into time-independent and time-dependent. 

\subsection{Time-independent Constraints}
This class of constraints determines the crew route among depots and compromised load buses and comprises
{
\small
\begin{subequations}\label{eq: RCR constraints 1}
\begin{align}
&\sum_{\forall i \in \mathcal{A}}X_{RC}^{(i,st,c)}=
\sum_{\forall j \in \mathcal{A}}X_{RC}^{(en,j,c)}=
\sum_{\forall i=j \in \mathcal{A}}X_{RC}^{(i,j,c)}=0, \label{Za} \\
&\sum_{\forall j \in \mathcal{A}/\{s,e\}}X_{RC}^{(st,j,c)}=
\sum_{\forall i \in \mathcal{A}/\{s,e\}}X_{RC}^{(i,en,c)}=1, \forall c\in \mathcal{C}. \label{Zb}\\
&\quad\sum_{\forall j \in \mathcal{A}}X_{RC}^{(i,ij,c)}=
\sum_{\forall i \in \mathcal{A}}X_{RC}^{(ij,j,c)}, \quad \forall ij \in \mathcal{A}/\{st,en\}, \label{Zc}\\
&\quad\sum_{\forall j \in \mathcal{A}}X_{RC}^{(i,j,c)} \le 1, \quad \forall i \in \mathcal{A}/\{en\}. \label{Zd}
\end{align}
\end{subequations}}

Constraints \eqref{Za} are to ensure that no repair crew returns back to the start depot, no crew leaves from the end depot, and no crew leaves from and returns back to the same bus. Constraints \eqref{Zb} make sure that each crew starts the route from depot $st$ and end at depot $en$. Constraints \eqref{Zc} describe that a crew arriving at a compromised load bus leaves it after finishing the malware removal job. Constraints \eqref{Zd} show that in each travel a crew can only visit one compromised load bus. To avoid the case that a crew first travels from buses $i$ to $j$, and then travels back from buses $j$ to $i$, which can make the solution useless, it is necessary to ensure
\begin{align}\label{eq: AndofBinaryVariables}
X_{RC}^{(i,j,c)}X_{RC}^{(j,i,c)}=0, \forall i,j\in\mathcal{A}.
\end{align}The product of two binary variables will induce non-linearity, and thus equations \eqref{eq: AndofBinaryVariables} are rewritten as
\begin{subequations}\label{eq: RCR constraint 2}
\begin{align}
    &X_{RC^*}^{(i,j,c)} \ge X_{RC}^{(i,j,c)}+X_{RC}^{(j,i,c)}-1, X_{RC^*}{(i,j,c)}=0, \\
    &X_{RC^*}^{(i,j,c)} \le X_{RC}^{(i,j,c)}, X_{RC^*}^{(i,j,c)} \le X_{RC}^{(j,i,c)},
    \forall i,j\in\mathcal{A}, c\in \mathcal{C}.
\end{align}
\end{subequations}

Moreover, each compromised load bus should be visited by one crew, which is represented by
\begin{align}
\sum_{\forall c\in \mathcal{C}}\sum_{j\in\mathcal{A}/\{st\}}X_{RC}^{(i,j,c)}=1, \forall i \in \mathcal{A}/\{en\}.
\end{align}

According to the repair crew route, the repair statuses of compromised buses can be calculated as
\begin{align}
Y_{RC}^{(i,c)}=\sum_{j\in\mathcal{A}/\{st\}}X_{RC}^{(i,j,c)}, \forall i \in \mathcal{A}/\{en\}.
\end{align}

\subsection{Time-dependent Constraints}
Based on the path travel and malware repair time, the time-dependent constraints are introduced to determine when the compromised buses are repaired and available. At first, the arrival time of repair crews at compromised buses are demonstrated as
\begin{subequations}
    \begin{align}
        &\quad\quad\quad AT_{RC}^{(st,c)}=0, \quad 0 \le AT_{RC}^{(i,c)} \le Y_{RC}^{(i,c)} \label{eq: AT1}\\
        &-B(1-X_{RC}^{(i,j,c)})\le AT_{RC}^{(i,c)}+R^{(i,c)}+T^{(i,j,c)}-AT_{RC}^{(j,c)} \nonumber \\
        &\quad \le B(1-X_{RC}^{(i,j,c)}), \quad\forall c\in \mathcal{C}, i,j\in\mathcal{A}/\{en\}, \label{eq: AT2}    \end{align}
\end{subequations}where constraints \eqref{eq: AT1} are to set the repair start time as $0$ and set the arrival time of the compromised buses that have not been visited by crews as $0$. Constraints \eqref{eq: AT2} are used to set the arrival time considering travel and repair time when the crew indeed moves along specific paths. For example, if $X_{RC}^{(i,j,c)}=1$, then $AT_{RC}^{(j,c)}$ will be equal to $AT_{RC}^{(i,c)}+R^{(i,c)}+T^{(i,j,c)}$. 

According to the arrival time at compromised buses, the repair status and availability can be then flagged by the following constraints
{\small
\begin{subequations}\label{eq: RCR constraint 7}
    \begin{align}
        &\sum_{\forall t \in \mathcal{T}}F_{RC}^{(i,t)}=1, \sum_{\forall c\in \mathcal{C}}(AT_{RC}^{(i,c)}+R_{i,c}Y_{RC}^{(i,c)}) \le \sum_{\forall t \in \mathcal{T}}tF_{RC}^{(i,t)}, \label{eq: F1} \\
        &\quad\sum_{\forall t \in \mathcal{T}}tF_{RC}^{(i,t)}\le \sum_{\forall c\in \mathcal{C}}(AT_{RC}^{(i,c)}+R_{i,c}Y_{RC}^{(i,c)})+1-\epsilon, \label{eq: F2} \\
        &Z_{RC}^{(i,0)}=0, Z_{RC}^{(i,t)}=\sum_{l\le t-1}F_{RC}^{(i,l)}, \label{eq: Z1} \quad\forall i \in\mathcal{A}/\{st,en\}, t\in\mathcal{T},
    \end{align}
\end{subequations}}where constraints \eqref{eq: F1} and \eqref{eq: F2} restrict that each compromised bus can only be repaired at one time step and the corresponding repair status is set strictly according to the arrival time and required repair time. More specifically, the small $\epsilon>0$ is introduced to allow to flag repair statuses at round time steps. For example, assume the repair crew arrive at bus $i$ at $t=1$, and the required repair time is $0.5$ time step, then its repair statuses over the time window with $5$ time steps are $\big\{F_{RC}^{(i,t)},t\in\{1,\cdots,5\}\big\}=\{0,1,0,0,0\}$. Equations \eqref{eq: Z1} set the availability of the compromised buses along with time variation under recovery planning. In the above example, the availability of bus $i$ over the time window should be $\big\{Z_{RC}^{(i,t)}\big\}=\{0,0,1,1,1\}$.

The description of RCR constraints ends here, and in the subsequent subsections, the availability variables will be used to select the eigenvalue sensitivity matrices that guide the DGA without violating the stability constraints. 

\section{Stability Constraints with Droop Gain Adjustment}\label{Sec: DGA constraints}
The DGAS constraints are to guarantee the frequency stability of \eqref{eq:frequency dynamics} under DLAAs, CRED, and recovery actions, which can be identified by its eigenvalues' real parts \cite{10124159}. Since it is difficult to give an explicit expression of the eigenvalue in terms of matrix entries, this section will first introduce the generation of eigenvalue sensitivity matrices, based on which the eigenvalue can be estimated accurately by strategically selecting the sensitivity information of sampling points, followed by the DGAS constraints.

\subsection{Eigenvalue Sensitivity Matrices}
Assume $K^{(i,j)}_{LL}=\widetilde{K}_{(i,j)}^{LL}=0,\forall i,j\in\mathcal{L}$, the dynamics \eqref{eq:frequency dynamics DLAA and Cyber Safe} can be rewritten as

\begin{align}\label{eq: transformed frequency dynamics}
\dot{\mathcal{X}}-
\mathcal{B}\mathcal{X}=\cb{\zeta},
\end{align}where $\mathcal{X}=[\cb{\delta}^{\rm T},\cb{\theta}^{\rm T},\cb{\omega}^{\rm T}]^{\rm T}$, $\mathcal{B}=$
\begin{align*}
&{\small
\begin{bmatrix}
I & \overleftrightarrow{0} & \overleftrightarrow{0} \\
\overleftrightarrow{0} & -(D_L)^{-1} & \overleftrightarrow{0} \\
\overleftrightarrow{0} & \overleftrightarrow{0} & -M \\
\end{bmatrix}
\begin{bmatrix}
\overleftrightarrow{0} & \overleftrightarrow{0} & I \\
H_{LG} & H_{LL} & -K_{LG}+\widetilde{K}_{LG}  \\
K_I+H_{GG} & H_{GL} & K_P+D_G \\
\end{bmatrix}},
\end{align*}and $\cb{\zeta}=(D_L)^{-1}(\cb{p}_{C*}-\cb{p}_{LS})[\overrightarrow{0},\overrightarrow{1},\overrightarrow{0}]^{\rm T}$.

The \textit{right} and \textit{left} eigenvalue problems associated with \eqref{eq: transformed frequency dynamics} can be expressed as
\begin{subequations}
    \begin{align}
        &\lambda_n\cb{r}_n-\mathcal{B}\cb{r}_n=\overrightarrow{0},  \label{eq: RightEigen}\\
        &\lambda_n\cb{l}_n^{\rm T}-\cb{l}_n^{\rm T}\mathcal{B}=\overrightarrow{0}, \forall n \in \Lambda. \label{eq: LeftEigen}
    \end{align}
\end{subequations}Differentiate both sides of equations \eqref{eq: RightEigen} with respect to $\widetilde{K}^{(i,s)}_{LG}$, and substitute \eqref{eq: LeftEigen} into the derived results, then the eigenvalue sensitivity can be obtained as
\begin{align}
    \frac{\partial \lambda_n}{\partial \widetilde{K}^{(i,s)}_{LG}}=-\frac{\cb{l}_n^{\rm T}\frac{\partial \mathcal{B}}{\partial \widetilde{K}_{LG}^{(i,s)}}\cb{r}_n}{\cb{l}_n^{\rm T}\cb{r}_n}, n \in \Lambda.
\end{align} 

Under the variation of multiple IBR droop gains, the eigenvalue can be estimated as
\begin{align}\label{eq: direct calculation without update}    \hat{\lambda}_n(\tilde{\cb{k}}_{LG})=\lambda_n|_{\tilde{\cb{k}}_{LG}=\overrightarrow{0}}+\Bigg(\frac{\partial \lambda_n}{\partial \tilde{\cb{k}}_{LG}}\Bigg|_{\tilde{\cb{k}}_{LG}=\overrightarrow{0}}\Bigg)^{\rm T}\tilde{\cb{k}}_{LG}.
\end{align}

However, the estimation will become inaccurate as the increase of $\tilde{\cb{k}}_{LG}$, since the adopted sensitivity information only uses the eigenvalue trend at the initial point. This inaccuracy can induce consequences like falsely assuming the unstable system as stable, which may invalidate the method in realistic applications.

\begin{figure}
\centering
\includegraphics[width=8cm]{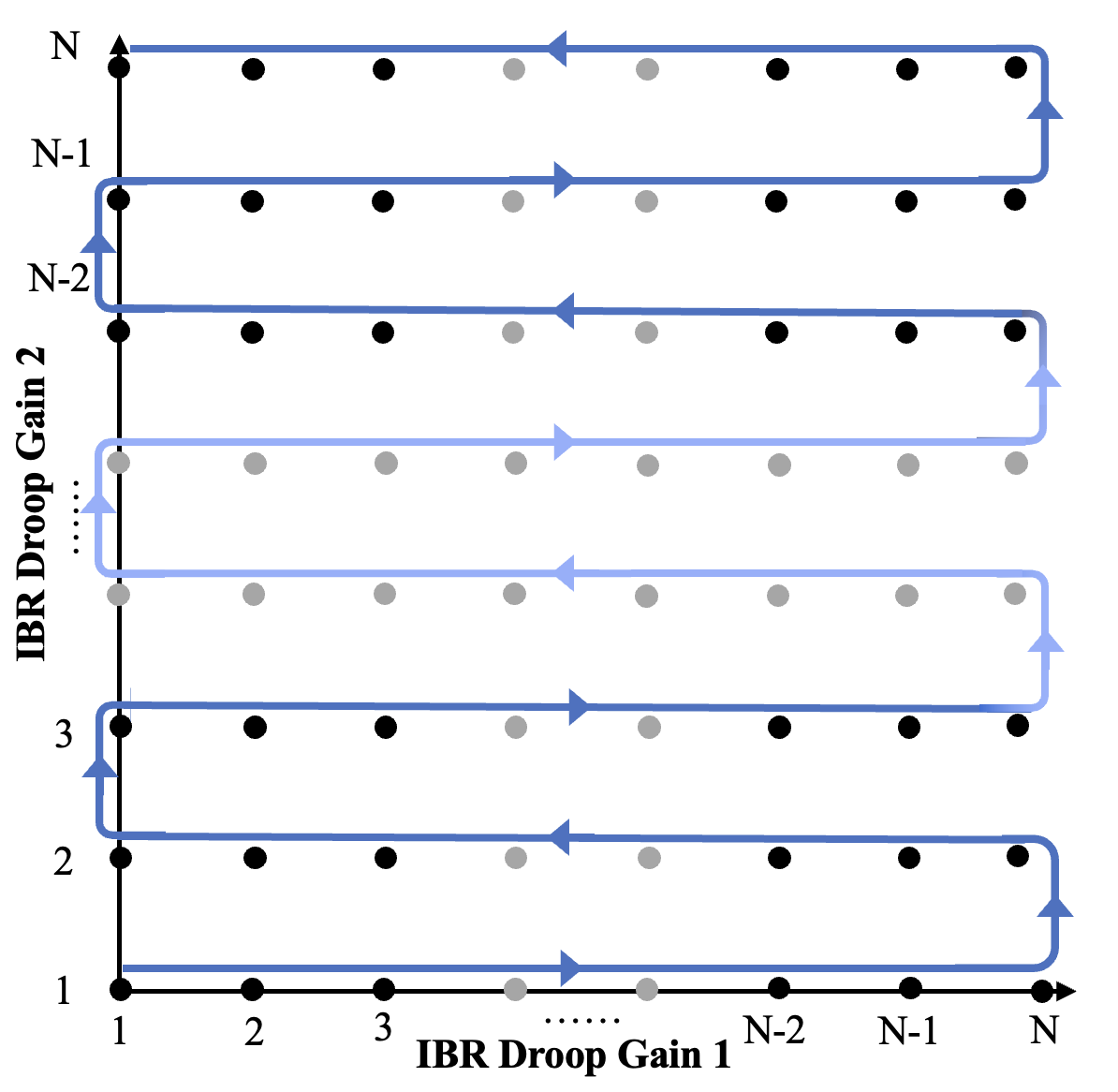}
\captionsetup{font={small}}
{\caption{Demonstration of the order of combinational sampling point, reflected by the arrow direction close to the solid points, with two IBR droop gains. The grey solid points indicate the omitted sampling points.}\label{fig: OrderofSamplingPointCombination}}
\end{figure}

Similar as \cite{chu2022mitigating}, we aim to incorporate the sensitivity information of multiple points strategically into the eigenvalue estimation to improve its accuracy. The basic idea is to update the selected start point $\tilde{\cb{k}}_{LG}$ and associated start eigenvalue $\lambda_n$ as well as eigenvalue sensitivity $\frac{\partial \lambda_n}{\partial \tilde{\cb{k}}_{LG}}$ used in \eqref{eq: direct calculation without update} if the eigenvalue estimation error exceeds the predefined threshold. But instead of considering only one IBR droop gain as in \cite{chu2022mitigating}, we need to estimate eigenvalues under the variation of \textit{multiple} IBR droop gains. In this case, the selection of sensitivity information is much more complicated due to the exponentially increasing search space. To reduce the recorded sensitivity information and simultaneously guarantee its estimation accuracy, all combinations of the sampling points indexing droop gains are carefully ordered. Assume that within the allowable ranges defined by equations \eqref{eq: IBR droop gain ranges1}-\eqref{eq: IBR droop gain ranges2} each $\widetilde{K}^{(i,s)}_{LG}$ is evenly sampled by $N$ points, and $N^{|\mathcal{D}|}$ combinations of sampling points that index all IBR droop gains are generated. These sampling point combinations are ordered in a consecutive manner such that any neighboring combinations only has one sampling interval difference on a IBR droop gain. Under this setting, the eigenvalue variation along this specific order is expected to be moderate over the evenly distributed alteration of IBR droop gains, and thus the sensitivity information extracted from one combinational sampling point can be used for the eigenvalue estimation at subsequent points as much as possible. A two-dimensional diagram is given in Fig. \ref{fig: OrderofSamplingPointCombination} to demonstrate the order of sampling point combinations with two IBR droop gains. 

Since the attacker can perceive the defense side's recovery actions, attack movements are incorporated into the eigenvalue sensitivity matrices. In particular, in each combination of compromised buses' availability, the worst case's attack parameters are used. Consider $|\mathcal{A}|-2$ compromised buses, there can be $2^{|\mathcal{A}|-2}$ combinations of compromised buses' availability along with the recovery process. Suppose that the start and end depots are put in the last two positions of $\mathcal{A}$, then we can use a natural number $m=\sum_{i=1}^{|\mathcal{A}|-2}(1-Z_{RC}^{(i)})2^{i-1}\in\mathcal{M}$ to index each combination. As it is difficult for the adversary to obtain the exact IBR droop gains in the time-critical recovery process, we consider the setting that the adversary has no knowledge of the IBR droop gains and will set them to zeros when designing attack strategies. To drive the dynamics \eqref{eq:frequency dynamics DLAA and Cyber Safe} to the unstable region as much as possible, the attack control gains are designed to maximize the real part of eigenvalues by solving the following problems:
\begin{subequations}
\begin{align}
    &\cb{k}_{LG}^{m*,n}=\arg\min_{\cb{k}_{LG}^m} -\Re(\lambda_n|_{\tilde{\cb{k}}_{LG}=\overrightarrow{0}}), \label{OP: optimal control gain1}  \\
    &\cb{k}_{LG}^{m*}=\arg\min_{\cb{k}_{LG}^{m*,n}}-\Re(\lambda_n^*|_{\tilde{\cb{k}}_{LG}=\overrightarrow{0}}), m \in \mathcal{M}, n \in \Lambda. \label{OP: optimal control gain2}
\end{align}
\end{subequations}Note that problems \eqref{OP: optimal control gain1}-\eqref{OP: optimal control gain2} can be formulated as an MILP problem using the sensitivity information with respect to attack control gains.
% Here we do not expand the details due to space limitation.

Based on the obtained worst-case attack gains $\cb{k}_{LG}^{m*}$, sensitivity matrices used for eigenvalue estimation under multiple IBR droop gains can be generated according to Algorithm \ref{Alg:sensitivity generation}. In particular, steps \ref{Order Matrix 1}-\ref{Order Matrix 2} obtain the order matrix $O\in\mathcal{R}^{|\mathcal{D}|\times N^{|\mathcal{D}|}}$ that sort all sampling point combinations of IBR droop gains in an expected sequence. Note that the details of functions {\sffamily repmat} and {\sffamily mod} can be found in the Matlab library. According to the allowable ranges of IBR droop gains, functions $\mathcal{F}(\cdot)$ are constructed to map the sampling indices to IBR droop gains. Then, initial values are given to the start eigenvalue, start point, and eigenvalue sensitivity through steps \ref{Sensitivty1}-\ref{Sensitivty2}. For each combination of IBR droop gain sampling points, the eigenvalue is first estimated using the sensitivity information at the last point through step \ref{Sensitivty5}, followed by the comparison with actual eigenvalue to decide if it is necessary to update the sensitivity information. In particular, if the estimation error is acceptable, then the sensitivity information will be copied from the last point using steps \ref{Sensitivty7}-\ref{Sensitivty8}. Otherwise, the sensitivity information will be updated at this combinational sampling point as in steps \ref{Sensitivty9}-\ref{Sensitivty10}.

\begin{algorithm}
\footnotesize
\caption{Eigenvalue Sensitivity Matrices Generation}\label{Alg:sensitivity generation}
\begin{algorithmic}[1]
\Require $N, \mathcal{D}, \mathcal{F}(\cdot), 
\cb{k}_{LG}^{m^*}, \lambda_n^{(j,m^*)}, \cb{l}_n^{(j,m^*)}, \cb{r}_n^{(j,m^*)}, \forall n \in \Lambda, m \in \mathcal{M}, j \in \mathcal{J};$
\Ensure $O,\lambda_{n,0}^{(j,m^*)}, \tilde{\cb{k}}_{LG,n,0}^{(j,m^*)}, \frac{\partial \lambda_n}{\partial \tilde{\cb{k}}_{LG}}\Big|_{(j,m^*)}, \forall n \in \Lambda, m \in \mathcal{M}, j \in \mathcal{J};$
\State \textbf{Obtain Order Matrix}\label{Order Matrix 1}
\State $\cb{v}_N=[1:N,N:-1:1];$
\For{$i=1:|\mathcal{D}|-1$}
\State $b_i=2N^i\lfloor \frac{N^{|\mathcal{D}|}}{2N^i} \rfloor$, $\cb{v}_N^{\otimes}=\cb{v}_N\otimes \overrightarrow{1}^{N^{(i-1)}}$;
\State $O(i,1:b_i)=\text{\sffamily repmat}(\cb{v}_N^{\otimes},1,\lfloor \frac{N^{|\mathcal{D}|}}{2N^i} \rfloor)$;
\If{$\text{\sffamily mod}(N^{|\mathcal{D}|},2N^i)$}
\State $O(i,b_i+1:\text{end})=[1:N]\otimes\overrightarrow{1}^{N^{(i-1)}}$;
\EndIf
\EndFor
\State $O(|\mathcal{D}|,:)=[1:N]\otimes\overrightarrow{1}^{N^{(|\mathcal{D}|-1)}}$;\label{Order Matrix 2}
\State \textbf{Calculate Sensitivity Matrix}
\State $\tilde{\cb{k}}_{LG}^{1}=\mathcal{F}\big(O(:,1)\big)$; 
% \Comment{Map sample indices to IBR gains}
\For{$m=1:2^{|\mathcal{A}|-2}$}
\For{$n=1:2|\mathcal{G}|+|\mathcal{L}|$}
\State
$\lambda_{n,0}^{(1,m^*)}=\lambda_n^{(1,m^*)}, \tilde{\cb{k}}_{LG,n,0}^{(1,m^*)}=\tilde{\cb{k}}_{LG}^{1}$;\label{Sensitivty1}
\State $\frac{\partial \lambda_n}{\partial \tilde{\cb{k}}_{LG}}\Big|_{(1,m^*)}=\frac{\partial \lambda_n}{\partial \tilde{\cb{k}}_{LG}}\Big|_{(\tilde{\cb{k}}_{LG}=\tilde{\cb{k}}_{LG}^{1},{\cb{k}}_{LG}={\cb{k}}_{LG}^{m^*})}$;\label{Sensitivty2}
\For{$j=2:N^{|\mathcal{D}|}$} \label{Sensitivty3}
\State $\tilde{\cb{k}}_{LG}^{j}=\mathcal{F}\big(O(:,j)\big)$;\label{Sensitivty4}
\State $\hat{\lambda}_{n}({\tilde{\cb{k}}_{LG}^{j}})=\hat{\lambda}_{n,0}^{(j-1,m^*)}+\Big(\frac{\partial \lambda_n}{\partial \tilde{\cb{k}}_{LG}}\Big|_{(j-1,m^*)}\Big)^{\rm T}\big(\tilde{\cb{k}}_{LG}^j-\tilde{\cb{k}}_{LG,n,0}^{(1,m^*)}\big);$\label{Sensitivty5}
\If{$\text{abs}(\hat{\lambda}_{n}({\tilde{\cb{k}}_{LG}^{j}})-\lambda_{n}^{(j,m^*)})\le \epsilon$}\label{Sensitivty6}
\State
{\small$\lambda_{n,0}^{(j,m^*)}=\lambda_{n,0}^{(j-1,m^*)}, \tilde{\cb{k}}_{LG,n,0}^{(j,m^*)}=\tilde{\cb{k}}_{LG,n,0}^{(j-1,m^*)}$};\label{Sensitivty7}
\State $\frac{\partial \lambda_n}{\partial \tilde{\cb{k}}_{LG}}\Big|_{(j,m^*)}=\frac{\partial \lambda_n}{\partial \tilde{\cb{k}}_{LG}}\Big|_{(j-1,m^*)}$;\label{Sensitivty8}
\Else
\State
$\lambda_{n,0}^{(j,m^*)}=\lambda_n^{(j,m^*)}, \tilde{\cb{k}}_{LG,n,0}^{(j,m^*)}=\tilde{\cb{k}}_{LG}^{j}$;\label{Sensitivty9}
\State $\frac{\partial \lambda_n}{\partial \tilde{\cb{k}}_{LG}}\Big|_{(j,m^*)}=\frac{\partial \lambda_n}{\partial \tilde{\cb{k}}_{LG}}\Big|_{(\tilde{\cb{k}}_{LG}=\tilde{\cb{k}}_{LG}^{j},{\cb{k}}_{LG}={\cb{k}}_{LG}^{m^*})}$;\label{Sensitivty10}
\EndIf
\EndFor
\EndFor
\EndFor
\end{algorithmic}
\end{algorithm}

% \begin{Remark}
% The core difference between the proposed eigenvalue estimation method and that of \cite{chu2022mitigating} is the need to address the simultaneous variation of multiple IBR droop gains. Since the relation between IBR droop gains and eigenvalues is highly nonlinear, it is not feasible to separately calculate the eigenvalue variation induced by each IBR droop gain, and then sum them up. Instead, the closely coupled eigenvalue impact caused by IBR droop gains should be taken into account. Therefore, besides updating the eigenvalue estimation related information along one nonzero IBR droop gain, more sampling points including multiple nonzero IBR droop gains should be checked for sensitivity information update. 
% \end{Remark}

\begin{figure}
    \centering
    \includegraphics[width=7.5cm]{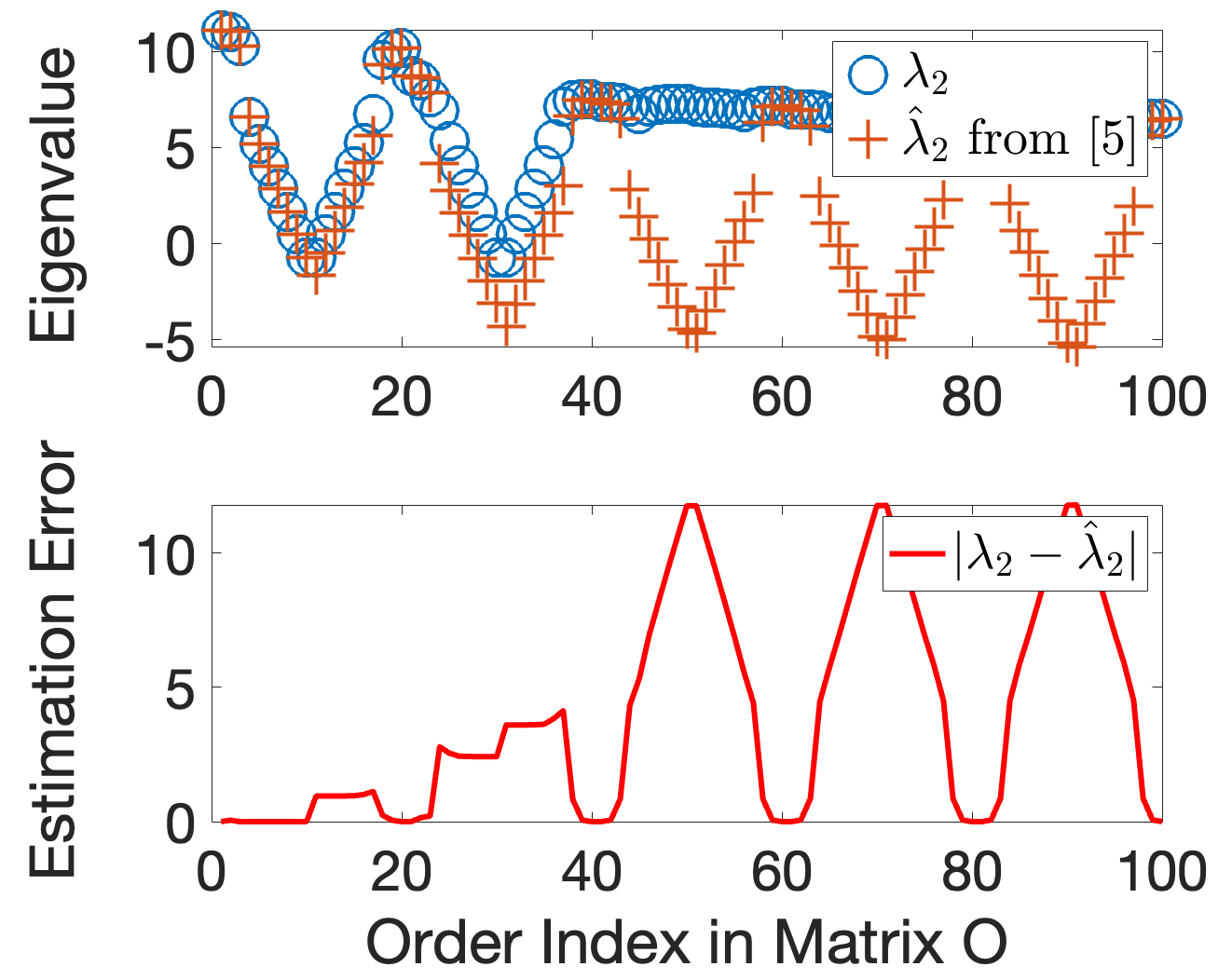}\\
    \includegraphics[width=7.5cm]{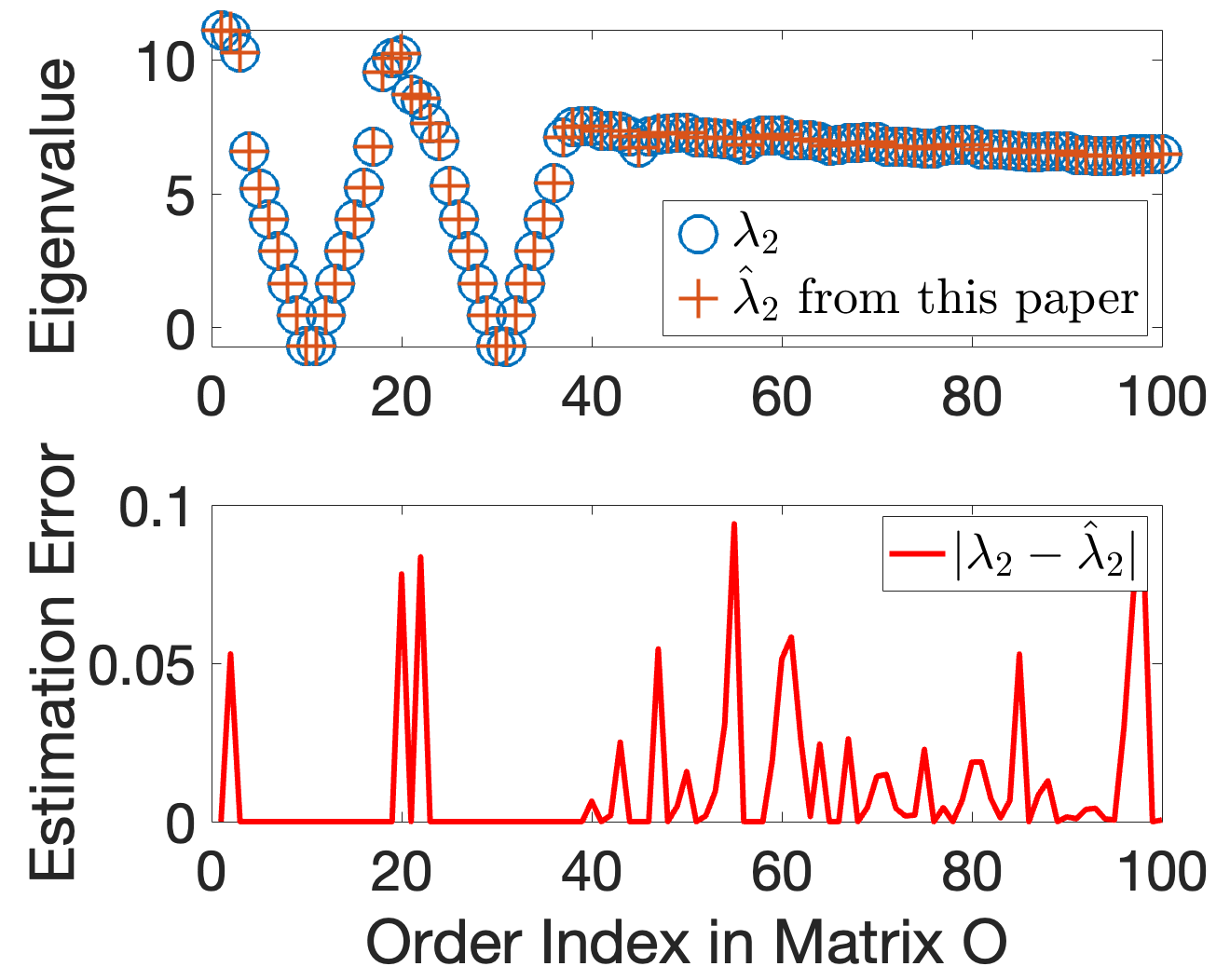}
    \caption{The upper sub-figure shows the accuracy of eigenvalue estimation using the method in \cite{chu2022mitigating}, while the lower sub-figure shows the accuracy of the proposed method.}\label{fig:EigenvaleEstimationComparsion}
\end{figure}

A simple example is given to illustrate the improved eigenvalue estimation accuracy. The attack control gains at $6$ compromised buses are set to their optimal values, under which the eigenvalue is estimated considering $2$ varying IBR droop gains. The number of sampling points is $N=10$. 
The results in Fig. \ref{fig:EigenvaleEstimationComparsion} indicate that the estimated eigenvalue $\hat{\lambda}_2$ using the method in \cite{chu2022mitigating} can deviate significantly from the actual one. An obvious consequence is that a large amount of unstable IBR droop gains are mistaken as stable. Oppositely, the proposed eigenvalue estimation method can bound the estimation error within $0.1$, customizable by the operator, as demonstrated in Fig. \ref{fig:EigenvaleEstimationComparsion}. The cost of this accuracy improvement is the higher computation and memory requirements in the proposed method, which is inevitable as the eigenvalue sensitivity matrices of more sampling points need to be computed and stored. In the above example, the method in \cite{chu2022mitigating} needs to compute and store $16$ sets of sensitivity matrices, while the proposed method requires $51$ sets of sensitivity matrices for the same case. Nevertheless, we believe that this potential issue can be well addressed in the near future with the rapid development of advanced digitised technologies like cloud computing and quantum computation.

% \begin{figure}
% \centering
% \begin{subfigure}[b]
% \includegraphics[width=9cm]{Picture/EigenvalueEstimationusingSingleValueVariation.png}
% \caption{ss1}
% \end{subfigure}
% \hfill
% \begin{subfigure}[b]
% \includegraphics[width=9cm]{Picture/EigenvalueEstimationusingMultipleValueVariation.png}
% \caption{ss}
% \end{subfigure}
% \caption{ss}
% \end{figure}

% \begin{figure}
%     \centering
%     \includegraphics[width=5cm]{Picture/EigenvalueEstimationusingSingleValueVariation.png}
%     \caption{This figure shows the accuracy of eigenvalue estimation using the proposed method in this paper.}\label{fig:EigenvaleEstimationSingle}
% \end{figure}

A binary variable $S_{DG}^{(m,t)}$ is introduced to select which set of eigenvalue sensitivity matrices should be used under the $m$-th recovery scenario of compromised buses. Corresponding constraints are given as
{\begin{subequations}\label{eq: DGA constraints1}
    \begin{align}
        &-(1-S_{DG}^{(m,t)})B\le\sum_{i=1}^{|\mathcal{A}|-2}(1-Z_{RC}^{(i,t)})2^{i-1}-m \label{Sequation1}, \\
        &\sum_{i=1}^{|\mathcal{A}|-2}(1-Z_{RC}^{(i,t)})2^{i-1}-m\le (1-S_{DG}^{(m,t)})B, \label{Sequation2} \\
        &\quad\quad\quad\quad\sum_{m\in\mathcal{M}} S_{DG}^{(m,t)}=1, \forall t \in \mathcal{T}, m\in\mathcal{M}, \label{Sequation3}
    \end{align}
\end{subequations}}where constraints \eqref{Sequation1}-\eqref{Sequation2} restrict that $S_{DG}^{(m,t)}$ can be either $0$ or $1$ if the calculated index at time $t$ is equal to $m$, and otherwise $S_{DG}^{(m,t)}$ will be set to $0$. With extra equations \eqref{Sequation3} requiring that at least one set of sensitivity matrices should be selected, $S_{DG}^{(m,t)}$ will be forced to be $1$ when $m$ matches the calculated index. Based on $S_{DG}^{(m,t)}$, we can choose the set of eigenvalue sensitivity matrices at time $t$, i.e.,
{\begin{subequations}\label{eq: DGA constraints2}
\begin{align}
\quad\quad\quad\quad\lambda_{n,0}^{(j,t)}&=\sum_{m\in\mathcal{M}}S_{DG}^{(m,t)}\lambda_{n,0}^{(j,m^*)}, \\
\quad\quad\quad\quad\tilde{\cb{k}}_{LG,n,0}^{(j,t)}&=\sum_{m\in\mathcal{M}}S_{DG}^{(m,t)}\tilde{\cb{k}}_{LG,n,0}^{(j,m^*)}, \\
\frac{\partial \lambda_n}{\partial \tilde{\cb{k}}_{LG}}\Big|_{(j,t)}&=\sum_{m\in\mathcal{M}}S_{DG}^{(m,t)}\frac{\partial \lambda_n}{\partial \tilde{\cb{k}}_{LG}}\Big|_{(j,m^*)},\\
&\forall j\in\mathcal{J}, t \in\mathcal{T}, n \in\Lambda \nonumber.
\end{align}
\end{subequations}}

\subsection{Stability Constraint Formulation}
This subsection explains how to formulate the stability constraints using the obtained eigenvalue sensitivity matrices. Let $\overline{\widetilde{K}^{(i,s)}_{LG}}$ and $\underline{\widetilde{K}^{(i,s)}_{LG}}$ be the upper and lower bounds of the $i$-th IBR droop gain, respectively, and $\xi_i=\frac{\overline{\widetilde{K}^{(i,s)}_{LG}}-\underline{\widetilde{K}^{(i,s)}_{LG}}}{N-1}$ be the sampling interval. Then the IBR droop gains of sampling points can be calculated as
\begin{align*}
\widetilde{K}_{LG}^{(i,s,l)}=\underline{\widetilde{K}^{(i,s)}_{LG}}+(l-1)\xi_i, \quad \forall l\in\{1,\cdots,N\}, i \in \mathcal{D}.
\end{align*}

For each sampling point, a range is defined to indicate that the IBR droop gain within this range should adopt this point's sensitivity matrices. The lower and upper bounds of these ranges are given as
\begin{align*}
    L^{(i,1)}=\widetilde{K}_{LG}^{(i,s,1)},\quad L^{(i,l)}=\widetilde{K}_{LG}^{(i,s,l)}-\frac{1}{2}\xi_i+\epsilon, 2\le l\le N, \\
    U^{(i,l)}=\widetilde{K}_{LG}^{(i,s,l)}+\frac{1}{2}\xi_i, 1\le l\le N-1,\quad U^{(i,N)}=\widetilde{K}_{LG}^{(i,s,N)},
\end{align*}$\forall i \in\mathcal{D}$, among which small $\epsilon>0$ is added to the lower bounds of ranges except the first one to avoid overlap of these ranges. To further decide which range a specific IBR droop gain belongs to, binary variables $T^{(i,l,t)}_{1}, T^{(i,l,t)}_{2}, T^{(i,l,t)}$ are introduced to satisfy
\begin{subequations}\label{eq: DGA constraints3}
    \begin{align}
        &\epsilon \le {L}^{(i,l)}-\widetilde{K}^{(i,s,t)}_{LG}+MT^{(i,l,t)}_{1}\le B,\\
        &\epsilon \le \widetilde{K}^{(i,s,t)}_{LG}-{U}^{(i,l)}+MT^{(i,l,t)}_{2}\le B,\\
        &T^{(i,l,t)}=T^{(i,l,t)}_{1}+T^{(i,l,t)}_{2}-1, \\
        &\forall i\in\mathcal{D}, 1\le l\le N, t\in\mathcal{T},
    \end{align}
\end{subequations}where $T^{(i,l,t)}$ will be set to $1$ only if ${L}^{(i,l)}\le\widetilde{K}^{(i,s,t)}_{LG}\le{U}^{(i,l)}$. 

Calculate the logic “and” of the position indicators of IBR droop gains corresponding to each column of matrix $O$, we can know which column the combination of IBR droop gains matches. The related constraints are written as
\begin{subequations}\label{eq: DGA constraints4}
    \begin{align}
        \Psi^{(j,t)}&\ge \sum_{i \in\mathcal{D}}T^{\big(i,O(i,j),t\big)}-\big(|\mathcal{D}|-1\big), \\
        \Psi^{(j,t)}\overrightarrow{1} &\le \Big[T^{\big(i,O(i,j),t\big)},\forall i \in\mathcal{D}\Big]_{|\mathcal{D}|}, \forall j\in\mathcal{J}, t \in \mathcal{T}.
    \end{align}
\end{subequations}

Then the estimated eigenvalue can be written as
\begin{subequations}
\begin{align}
\hat{\lambda}_{n,t}=\sum_{j\in\mathcal{J}}\Psi^{(j,t)}&\Bigg[\lambda_{n,0}^{(j,t)}+\Bigg(\frac{\partial \lambda_n}{\partial \tilde{\cb{k}}_{LG}}\Big|_{(j,t)}\Bigg)^{\rm T}*\\
&\Big(\tilde{\cb{k}}_{LG}^t-\tilde{\cb{k}}_{LG,n,0}^{(j,t)}\Big)\Bigg], \forall  n \in \Lambda, t \in \mathcal{T},
\end{align}
\end{subequations}and the stability constraints are
\begin{align}\label{eq: DGA constraints6}
\Re\big(\hat{\lambda}^{(n,t)}\big)\le -\epsilon, \quad \forall  n \in \Lambda, t \in \mathcal{T}.
\end{align}

\begin{figure}
    \centering
    \includegraphics[width=8cm]{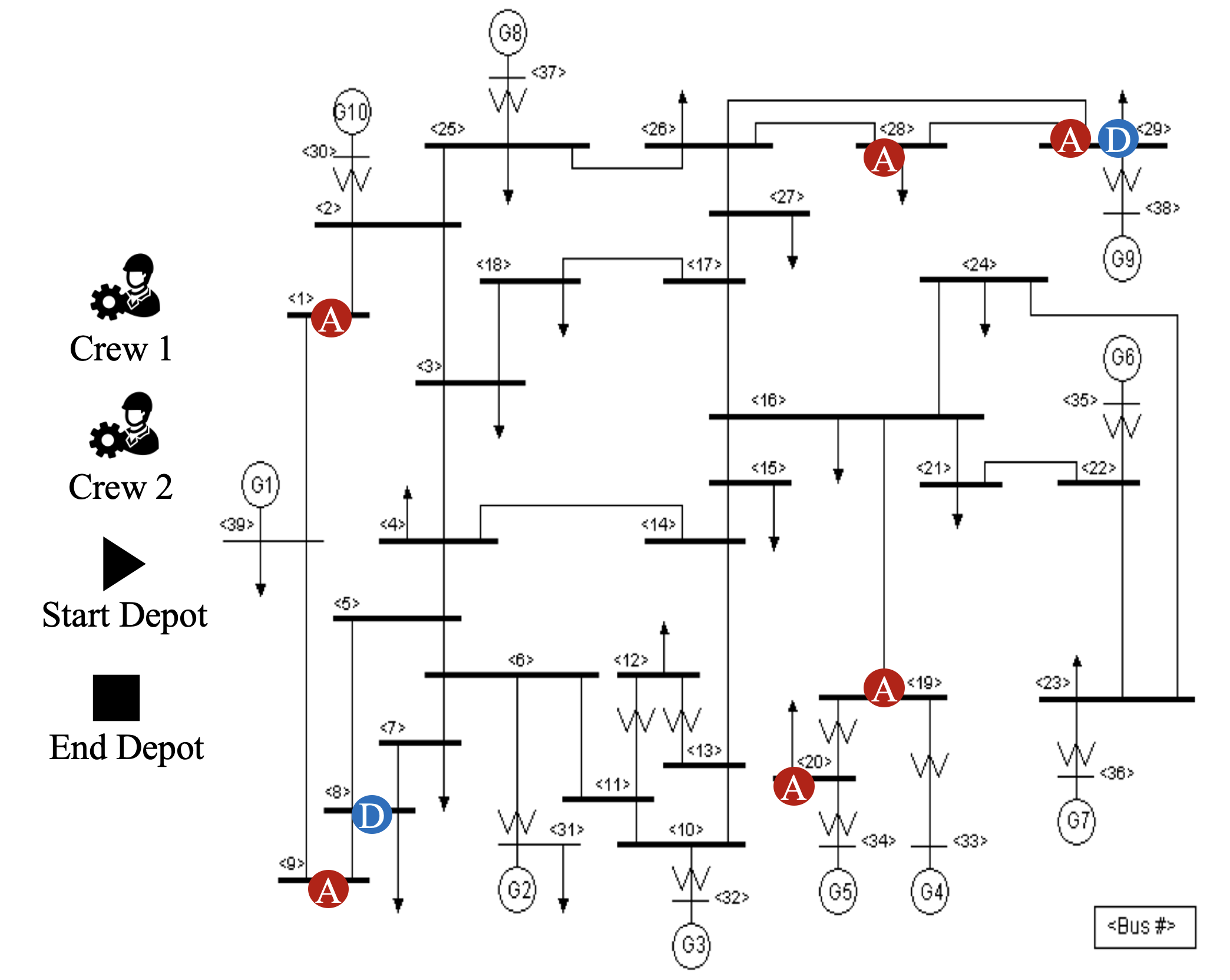}
    \captionsetup{font={small}}
    \caption{This figure shows the modified IEEE 39-bus power system, where red solid circles with letter "A" mean that some IoT-based loads connected to these buses are compromised, and blue solid circles with letter $D$ indicate that IBRs connected to these buses are used for CRED. Besides, repair crews, start, and end depots are also depicted in the left.}\label{fig:IEEE39buscase}
\end{figure}

\section{Case Studies} \label{section: case studies}
In this section, the effectiveness of the proposed CRDA is assessed in the modified IEEE 39-bus power system case as shown in Fig. \ref{fig:IEEE39buscase}. The set of load buses is $\mathcal{L}=\{1,\cdots,29\}$ and the set of generator buses is denoted by $\mathcal{G}=\{30,\cdots,39\}$. The parameters of transmission lines and the inertial and damping parameters are as in \cite{pai1989energy,athay1979practical}. Synchronous generator controller parameters are $\cb{k}_P=[10,4.5,4.5,1,5,4,3,2,4,5]$ and $K_I^{(i)}=6,\forall i \in\mathcal{G}$. The damping parameter of each frequency-sensitive load is $1$. Vulnerable load buses, maximal vulnerable loads, and their repair time are shown in TABLE \ref{tab:Vulnerable bus indices}. Assume the safety limitation of generator tripping $\omega_s^{max}=\frac{2}{50}$ p.u., and then the maximal allowable attack control gains can be calculated according to \eqref{eq: attack control gain range} as $\cb{k}_{KG}\le[11,9,14,10,12,9]^{\rm T}$. Moreover, IBRs connected to buses $\{8,29\}$ are used to mitigate the stability impact induced by DLAAs, with the frequency measurements from buses $\{39,38\}$. The maximal allowable IBR droop control gains are $\widetilde{K}_{LG}^{(8,39)}, \widetilde{K}_{LG}^{(29,38)}\in[0,15]$. Through numerical studies, we notice that the optimal (worst-case) attack control gains $\cb{k}_{LG}^{m^*}$ are very close to their allowable upper bounds under all $2^6$ attack scenarios, and thus here we do not list these parameters separately. By appropriately adjusting the IBR droop control gains, the destabilizing impact of DLAA can be effectively mitigated as in \cite{chu2022mitigating}. Two repair crews are considered in the case studies, and in general the repair time depends on the amount of compromised loads. However, as the type of compromised IoT-based load (like Android- or Linux-based) differs and the expertise of each crew varies, the repair time against these loads can have slight discrepancy from that based on load amounts. The weight parameters of attack recovery are set according to the amount of compromised loads as $\cb{\beta}=0.01*[0.84,0.80,0.70,0.80,0.80,1.00]^{\rm T}$. The number of sampling points for each IBR droop gain $N=4$, the Big-M penalty parameter $B=1e4$, and the small parameter $\epsilon=1e-4$. Moreover, the recovery plan is scheduled over $20$ time steps, with each step representing $30$ minutes.

\begin{table}[h]
\begin{center}
\footnotesize
\centering
\captionsetup{font={small}}
\caption{Vulnerable Bus Index, Maximal Vulnerable Load, and Repair Time}\label{tab:Vulnerable bus indices}
\begin{tabular}{ p{1cm} <{\centering}| p{1.5cm} <{\centering} | p{2cm} <{\centering} | p{2cm} <{\centering}}
\toprule[1pt]
Index & \multirow{2}{2cm}{$\overline{P}_{LV}^{(i)}$ [p.u.]} & \multicolumn{2}{c}{Repair time (30 minutes step)} \\ 
$(i,s)$ & & Crew 1 & Crew 2 \\ \hline
(1,39) & 0.84 & 3.50 & 3.50  \\
(9,39) & 0.80 & 3.00 & 3.50 \\
(19,33) & 0.70 & 3.00 & 4.00 \\
(20,34) & 0.80 & 3.50 & 4.50 \\
(28,38) & 0.80 & 3.50 & 3.00 \\
(29,38) & 1.00 & 4.00 & 4.50 \\
\bottomrule[1pt]
\end{tabular}
\end{center}
\end{table}
% \vspace{-30pt}
\subsection{Effectiveness of the Proposed CRDA}
To illustrate the proposed CRDA's effectiveness, we compare its result with that of a benchmark method, where the RCR and DGA problems are decoupled and solved separately. In the first stage, the RCR problem is solved using 
{\begin{subequations}\label{eq: SP objective function 1}
\begin{align}
    &\min_{\Gamma} -\sum_{\forall t\in\mathcal{T}}\sum_{\forall i \in \mathcal{A}/\{st,en\}}\beta_i Z_{RC}^{(i,t)}, \\
    &\text{s.t.\quad RCR constraints: } \eqref{eq: RCR constraints 1}, \eqref{eq: RCR constraint 2}-\eqref{eq: RCR constraint 7}.
\end{align}
\end{subequations}}
Based on the RCR solution $\Gamma^*$, the optimal attack control gains $\cb{k}_{LG}^{m^*}$ are calculated and corresponding eigenvalue sensitivity matrices are obtained using Algorithm \ref{Alg:sensitivity generation}. Then, the smallest IBR droop gains are calculated to stabilize the frequency under DLAAs, and then the DGA problem is solved, which can be formally formulated as

{\begin{subequations}\label{eq: SP objective function 2}
\begin{align}
    &\quad\quad\min_{\Upsilon} \quad \sum_{\forall t\in\mathcal{T}}\sum_{\forall i\in\mathcal{D}}\widetilde{K}_{LG}^{(i,s,t)}, \\
    &\text{s.t. \ DGAS constraints: } \eqref{eq: DGA constraints1}, \eqref{eq: DGA constraints2}, \eqref{eq: DGA constraints3}-\eqref{eq: DGA constraints6}.
\end{align}
\end{subequations}}

\begin{table}[h]
\begin{center}
\footnotesize
\centering
\captionsetup{font={small}}
\caption{RCRs obtained from the RCR-DGA Decoupled and proposed CRDA Problems}
\label{tab:RCRsundertwomethods}
\begin{tabular}{ p{2cm} <{\centering}| p{6cm} <{\centering} }
\toprule[1pt]
 RCR-DGA & $\text{Crew}\ 1: st \rightarrow 29 \rightarrow 9 \rightarrow 19 \rightarrow en$ \\ 
 Decoupled & $\text{Crew}\ 2: st \rightarrow 28 \rightarrow 1 \rightarrow 20 \rightarrow en$ \\ \hline
 \multirow{2}{0.5cm}{CRDA} & $\text{Crew}\ 1: st \rightarrow 9 \rightarrow 29 \rightarrow 19 \rightarrow en$ \\ 
& $\text{Crew}\ 2: st \rightarrow 28 \rightarrow 1 \rightarrow 20 \rightarrow en$ \\

\bottomrule[1pt]
\end{tabular}
\end{center}
\end{table}

\begin{figure}
    \centering
    \includegraphics[width=7.5cm]{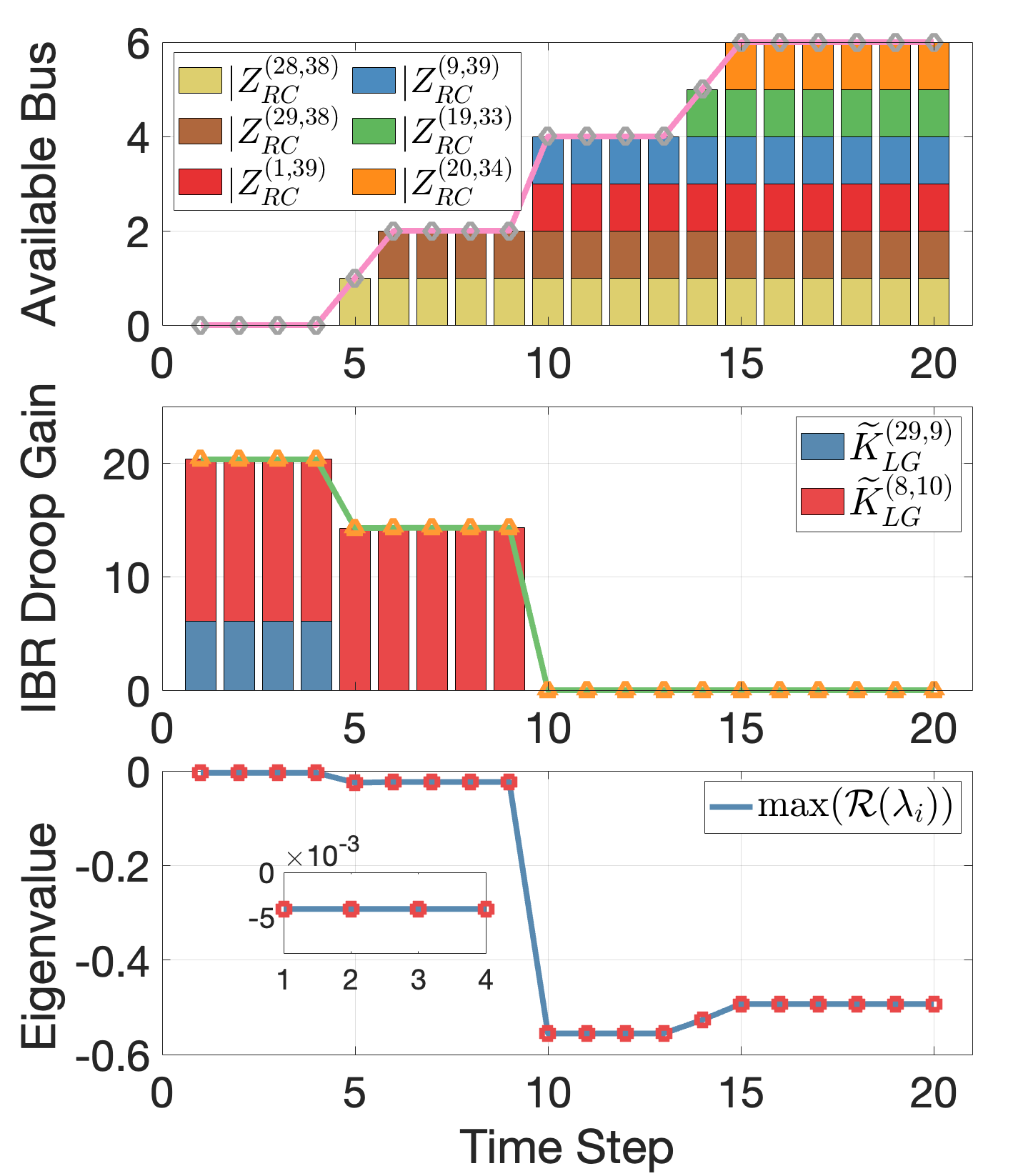}\\
    \includegraphics[width=7.5cm]{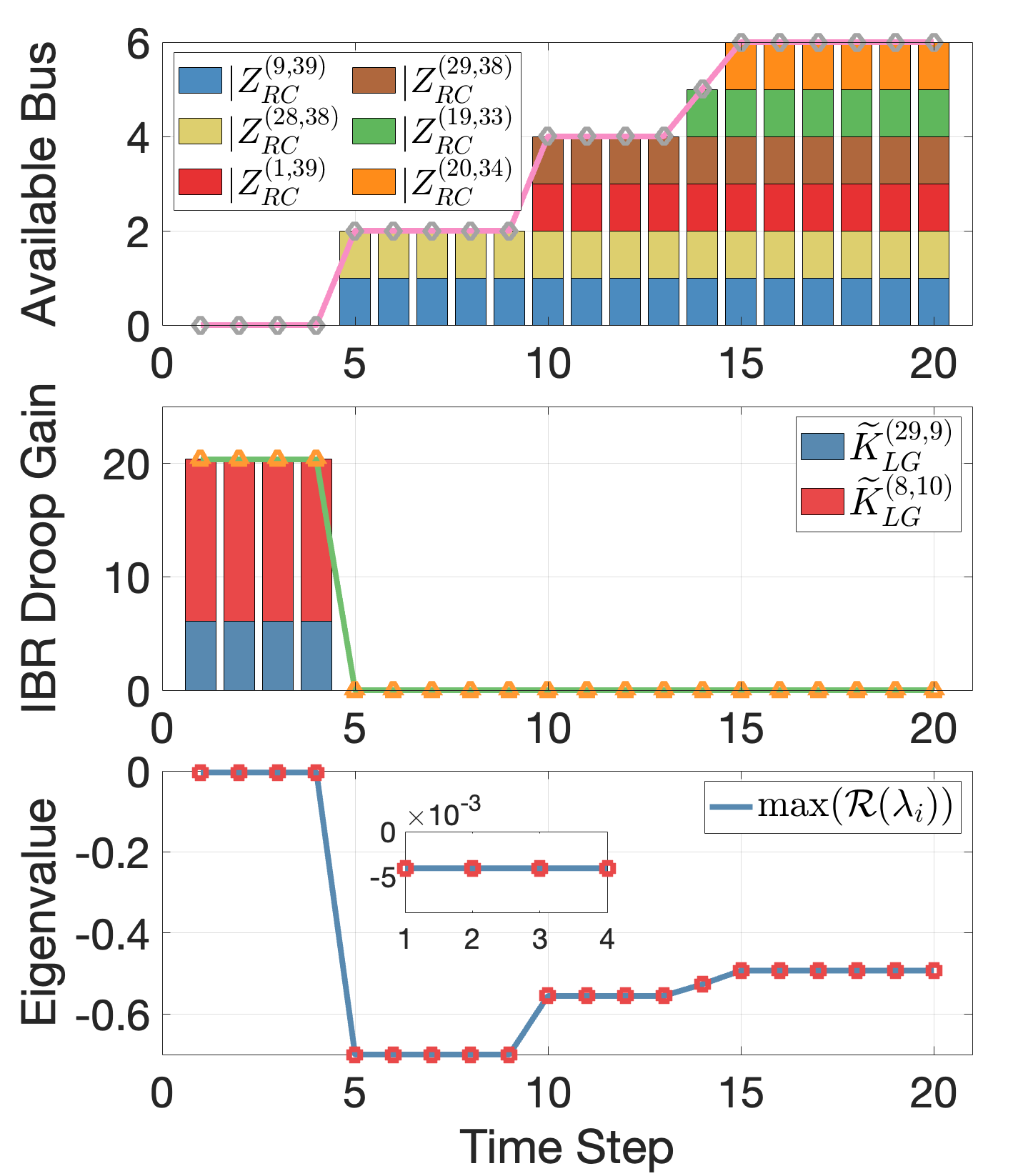}
    \captionsetup{font={small}}
    \caption{The upper sub-figure shows the results of the RCR-DGA decoupled method (benchmark) \eqref{eq: SP objective function 1}-\eqref{eq: SP objective function 2}, and the lower sub-figure shows those of the proposed CPRP under DLAAs \eqref{eq: overall objective function}. From top to down in each sub-figure, the results includ the bus availability, IBR droop gains, and eigenvalues over all time steps.}\label{fig:Effectiveness Comparsion Results}
\end{figure}

By solving the CRDA \eqref{eq: overall objective function} and RCR-DGA decoupled \eqref{eq: SP objective function 1}-\eqref{eq: SP objective function 2} problems, two RCRs are obtained as shown in TABLE \ref{tab:RCRsundertwomethods}. There is only one difference in the repair route for crew $1$, i.e., the repair order of buses $9$ and $29$. In the 
RCR-DGA decoupled problem, bus $29$ will be first repaired since the compromised load there is larger than that of bus $9$. While in the proposed CRDA problem, it is the opposite result as repairing the compromised load at bus $9$ is more beneficial for the DGA. This benefit is vividly demonstrated through Fig. \ref{fig:Effectiveness Comparsion Results}. Using the RCR obtained from the proposed CRDA problem, all IBR droop gains can be reset to zero at time step $5$. Nevertheless, the RCR under the RCR-DGA decoupled case requires $5$ more steps to reset all IBR droop gains. For illustrative purpose, assume that the operation cost of the IBR droop control is $254 \pounds/\widetilde{K}_{LG}^{(i)}=1/\text{time step}$ \cite{chu2022mitigating}, the proposed CRDA can save about $1.81 \text{k}\pounds $ operation cost compared with the RCR-DGA decoupled case. In general, the proposed CPRP can achieve a much quicker recovery from the CRED to the normal operational mode compared with the RCR-DGA decoupled process, while guaranteeing frequency stability and keeping the same recovery time for all compromised loads.

% \vspace{-10pt}

\subsection{Sensitivity of Optimality and Computation Time with respect to Sampling Number}
In this subsection, we assess the impact of sampling number $N$ on the optimal objective values and computation time. As shown in TABLE \ref{tab:Opsensitivity}, with an increased $N$, the computation time increases extremely with an exponential rate. However, the optimal objective values do not decrease accordingly as expected. The objective values under the case of $N=8$ even have slight (negligible) increase compared with those of previous cases. Therefore, although choosing a larger $N$ can in general lead to improved objective values, it is necessary to consider the associated cost of computation resource and set an appropriate $N$ to balance the trade-off between them. In this case study, it is verified that the setting of $N=4$ is cost-efficient.

\begin{table}[h]
\footnotesize
\begin{center}
\centering
\captionsetup{font={small}}
\caption{Optimality and Computation Time under Different Sampling Number $N$}\label{tab:Opsensitivity}
\begin{tabular}{ p{2.5cm} <{\centering}| p{1cm} <{\centering} | p{1cm} <{\centering} | p{1cm} <{\centering} }
\toprule[1pt]
$N$ & 4 & 6 & 8  \\ \hline
Overall Obj. & 81.02 & 80.89 & 80.91 \\
DGA Obj.  & 81.37 & 81.24 & 81.25\\
RCR Obj  & -0.35 & -0.35 & -0.34 \\
Comp. Time [s] & 193.47 & 322.94 & 1451.14 \\
\bottomrule[1pt]
\end{tabular}
\end{center}
\end{table}
% \vspace{-30pt}

\subsection{Discussion on the Scalability}
\begin{figure}
    \centering
    \includegraphics[width=6cm]{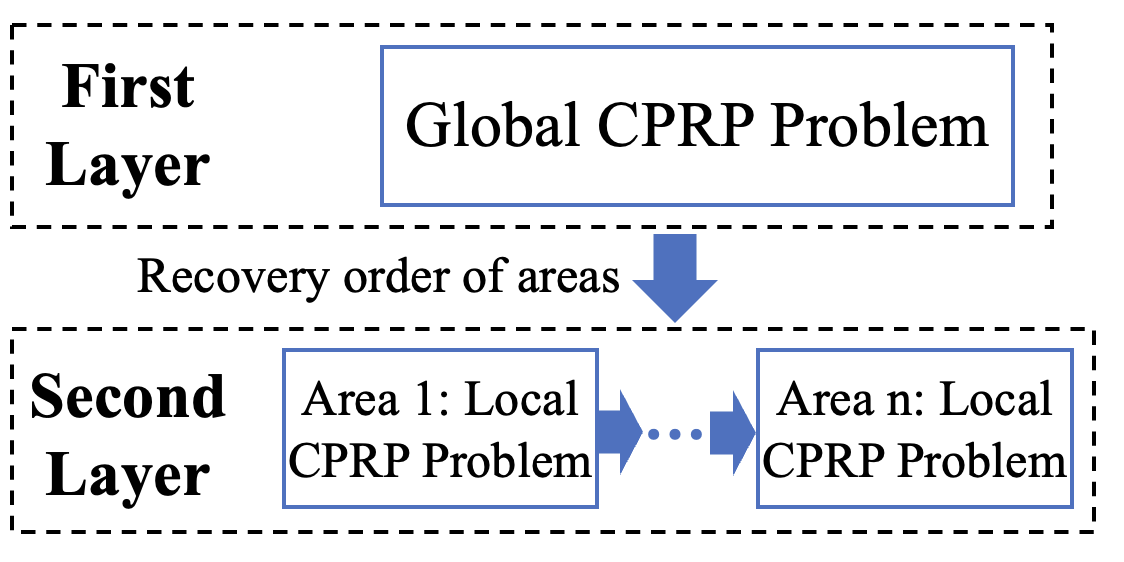}
    \captionsetup{font={small}}
    \caption{This figure shows the conceptual diagram of the hierarchical CRDA framework.}\label{fig:HierarchicalRecoveryPlanning}
\end{figure}

In this subsection, we discuss a potential hierarchical framework that can address the scalability issue of the proposed CRDA. Normally, the computation time can be extremely large (order of hours/days) when the number of buses equipped with IBRs is more than $10$. The essential idea to address this issue is to design the recovery planning in multiple layers. Consider the power system comprising multiple areas, and in each area, the generator inertial, damping, and IBR outputs are aggregated. In the first layer, a \textit{global} CRDA problem \eqref{eq: overall objective function} is formulated and solved from the area-based perspective, where each area is regarded as an aggregated bus. Note that in the first layer's problem, there should be only one repair team, i.e., $|\mathcal{C}^f|=1$, to ensure that the repair team is only within one area at a time such that the CRDA problems formulated for areas can be decoupled. In the second layer, following the recovery order of areas obtained from the first layer, each area will use a local CRDA problem to repair the compromised load buses inside it. A demonstrative diagram is depicted in Fig. \ref{fig:HierarchicalRecoveryPlanning} to illustrate the hierarchical CRDA framework.

% \vspace{-15pt}

\section{Summary and Future Directions}\label{Sec: Summary}
In this paper, we investigated the CRDA for the first time and formulated it as an MILP problem. In particular, RCR and DGAS constraints are formulated and incorporated to facilitate the fast restoration of the post-attack system to the normal operational mode without violating stability requirements. Sensitivity information of strategically selected sampling points are adopted to ensure the linearity of stability constraints, and attack movements are also incorporated using the worst case's attack parameters in each step. Lastly, case studies were conducted in the modified IEEE 39-bus power system to verify the improved performance of the proposed CRDA compared with the benchmark case where the RCR and DGA sub-tasks are considered separately. Furthermore, the sensitivity of computation time and result optimality with respect to the sampling number were illustrated and the potential scalability solution was discussed. Future directions include i) Applying machine learning methods to address the trade-off between computation time and result optimality and ii) Using gaming methods to model the competing process between attack and defense players.

\begin{spacing}{0.93}
% \balance
\bibliographystyle{IEEEtran}
\tiny
\bibliography{root}
\end{spacing}
\end{spacing}

\end{document}